\documentclass[aps,prxquantum,reprint,twocolumn,superscriptaddress,floatfix,nofootinbib,longbibliography]{revtex4-1}
\usepackage{graphicx,amsmath,amsfonts,amssymb,amsthm,xr}
\usepackage{epsfig,amsmath,amssymb,color,dsfont,upgreek,physics}
\usepackage{mathrsfs}
\usepackage{mathtools}
\usepackage{bbold}
\usepackage{comment}
\usepackage{float}

\usepackage[bookmarks=true,colorlinks,linkcolor=OrangeRed,urlcolor=NavyBlue,citecolor=RoyalBlue]{hyperref}
\usepackage[dvipsnames]{xcolor}
\usepackage{orcidlink}

\definecolor{mygold}{rgb}{0.5,0.6,0.7}

\definecolor{mypurple}{rgb}{0.49,0.18,0.56}

\definecolor{mygreen}{rgb}{0,0.5,0}
\definecolor{myred}{rgb}{0.7,0,0}
\definecolor{myblue}{rgb}{0,0,1}

\begin{document}
\title{A Cold-Atom Particle Collider}

\author{Guo-Xian Su${}^{\orcidlink{0000-0001-7936-762X}}$}
\thanks{These authors contributed equally to this work.}
\affiliation{Department of Modern Physics, University of Science and Technology of China, Hefei, Anhui 230026, China}
\affiliation{Physikalisches Institut, Ruprecht-Karls-Universit\"at Heidelberg, Im Neuenheimer Feld 226, 69120 Heidelberg, Germany}
\affiliation{CAS Center for Excellence and Synergetic Innovation Center in Quantum Information and Quantum Physics, University of Science and Technology of China, Hefei, Anhui 230026, China}

\author{Jesse Osborne${}^{\orcidlink{0000-0003-0415-0690}}$}
\thanks{These authors contributed equally to this work.}
\affiliation{School of Mathematics and Physics, The University of Queensland, St.~Lucia, QLD 4072, Australia}

\author{Jad C.~Halimeh${}^{\orcidlink{0000-0002-0659-7990}}$}
\email{jad.halimeh@physik.lmu.de}
\affiliation{Department of Physics and Arnold Sommerfeld Center for Theoretical Physics (ASC), Ludwig-Maximilians-Universit\"at M\"unchen, Theresienstra\ss e 37, D-80333 M\"unchen, Germany}
\affiliation{Munich Center for Quantum Science and Technology (MCQST), Schellingstra\ss e 4, D-80799 M\"unchen, Germany}
\affiliation{Dahlem Center for Complex Quantum Systems, Freie Universit\"at Berlin, 14195 Berlin, Germany}

\begin{abstract}
A major objective of the strong ongoing drive to realize quantum simulators of gauge theories is achieving the capability to probe collider-relevant physics on them. In this regard, a highly pertinent and sought-after application is the controlled collisions of elementary and composite particles, as well as the scattering processes in their wake. Here, we propose particle-collision experiments in a cold-atom quantum simulator for a $1+1$D $\mathrm{U}(1)$ lattice gauge theory with a tunable topological $\theta$-term, where we demonstrate an experimentally feasible protocol to impart momenta to elementary (anti)particles and their meson composites. We numerically benchmark the collisions of moving wave packets for both elementary and composite particles, uncovering a plethora of rich phenomena, such as oscillatory string dynamics in the wake of elementary (anti)particle collisions due to confinement. We also probe string inversion and entropy production processes across Coleman's phase transition through far-from-equilibrium quenches. We further demonstrate how collisions of composite particles unveil their internal structure. Our work paves the way towards the experimental investigation of collision dynamics in state-of-the-art quantum simulators of gauge theories, and sets the stage for microscopic understanding of collider-relevant physics in these platforms.
\end{abstract}

\date{\today} 
\maketitle
{\hypersetup{linkcolor=mygold}
\tableofcontents}

\begin{figure*}
	\centering
	\includegraphics[width=0.9\linewidth]{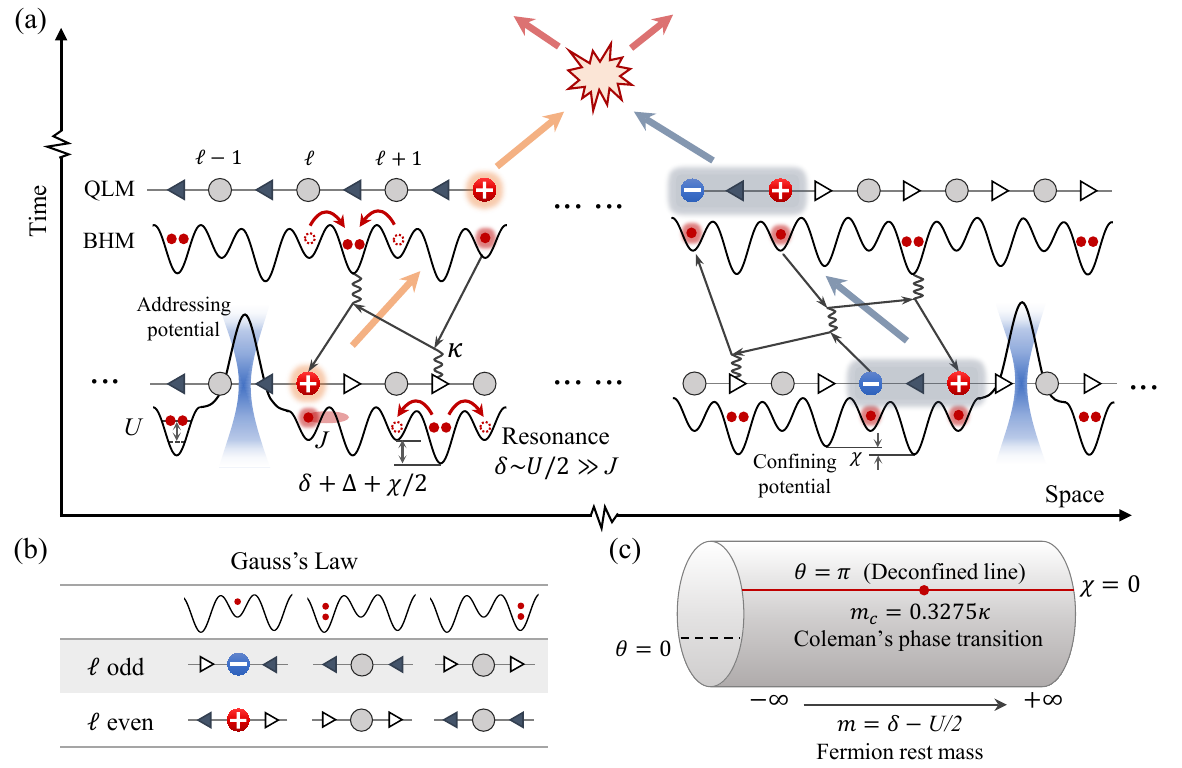}
	\caption{(Color online).
        \textbf{Schematics of a cold-atom particle collider.}
        (a) Implementing a $\mathrm{U}(1)$ lattice gauge theory in a Bose--Hubbard quantum simulator to explore collision dynamics. Bosons in the optical lattice can be described by the Bose--Hubbard model (BHM) with tunneling $J$, onsite interaction $U$, and chemical potential $\mu$, which is generated by the staggered optical superlattices with a period-$2$ staggering $\delta$, a period-$4$ staggering $\chi$, and a linear tilt $\Delta$. The gauge invariant coupling is realized by the second-order correlated hopping in BHM, with strength $\kappa \approx 8\sqrt{2}J^2/U$, and the (anti)particle rest mass $m=\delta-U/2$. A single (anti)particle excitation in the gauge theory is represented by a single boson on the (even) odd matter site. In the large mass limit, the hopping of an (anti)particle is a second-order process in the gauge theory with strength $\Tilde{t} \approx \kappa^2/8m$. To explore particle collisions in the quantum simulator, we prepare the uni-directional moving wave packets of a single (anti)particle or a composite particle (meson) by placing a potential barrier next to the particle, which can be realized in the experiment with a blue-detuned addressing beam. 
        (b) The local $\mathrm{U}(1)$ gauge symmetry manifests Gauss's law which only allows the configurations shown in the table when working in the physical gauge sector $\hat{G}_\ell \ket{\Psi}=0$. We use the staggered-fermion representation where particles and antiparticles reside on alternating matter sites.
        (c) The $\theta-m$ phase diagram of quantum electrodynamics. The staggering potential $\chi$ on gauge sites tunes the topological $\theta$-angle away from $\theta=\pi$, breaking the degeneracy between the two electric fluxes, which leads to confinement. At $\theta=\pi$, we have Coleman's phase transition at critical mass $m_\text{c}=0.3275\kappa$.
        }
	\label{fig:kick}
\end{figure*}

\section{Introduction}

Particle collider experiments are key to unlocking the nature of elementary particles and their interactions, and have yielded deep insights into the Standard Model of Particle Physics \cite{Ellis_book}. They unravel subatomic structures, enable the discovery of new particles \cite{ATLAS2012,CMS2012}, and allow the creation of quark-gluon plasmas that mimic the conditions of early universe cosmology \cite{Adcox2005, Back2005, Arsene2005}. Remarkably, with particle colliders, researchers are reaching towards physics even beyond the Standard Model, such as the Future Circular Collider (FCC) at CERN \cite{FCC}, which will search for evidence of the dark matter particles by the name of weakly interacting massive particles (WIMPs) \cite{Arcadi_review}.

The connection between theoretical predictions and observations in collision experiments currently relies heavily on numerical simulations \cite{Sjostrand1994}. Due to the highly nonperturbative and quantum many-body nature of various high-energy scattering events, there is no general \textit{ab initio} method on classical computers that can simulate their real-time collision dynamics from the far-from-equilibrium early stages to late-time equilibration. Among the most prominent classical methods, the highly successful quantum Monte Carlo simulations of lattice quantum chromodynamics (QCD) suffer from the sign problem in the application to real-time dynamics. Dedicated time-evolution methods, such as the time-dependent density matrix renormalization group (DMRG) method \cite{White2004,Schollwoeck_review,Uli_review,Paeckel_review}, are mostly restricted to spatially (quasi-)one-dimensional systems and to relatively short evolution times due to many-body entanglement buildup, which classical computers fundamentally cannot handle as the computational cost becomes exponential in available resources. As such, phenomenological models have traditionally been employed on classical computers to analyze collider data in order to better understand the underlying highly nonperturbative far-from-equilibrium phenomena \cite{Andersson1983}. However, these models are not exact, and they rely on various approximations. It is thus useful to seek alternate venues where such collider phenomena can be understood from first-principles time evolution, and where entanglement buildup can be naturally handled.

Inspired by Feynman's vision of simulating the dynamics of a quantum many-body system with an engineered tunable quantum simulator \cite{Feynman1982,Lloyd1996,Bloch2008,Hauke2012,Georgescu_review}, the application of quantum simulation to high-energy physics problems has made notable progress over the past years, with experimental demonstrations using trapped-ion platforms, superconducting qubits, and cold-atom quantum gases \cite{Dalmonte_review,Pasquans_review,Zohar_review,Alexeev_review,aidelsburger2021cold,zohar2021quantum,klco2021standard,Bauer_review,Bauer_ShortReview,funcke2023review,dimeglio_review,Halimeh_review,Martinez2016,Bernien2017,Dai2017,Klco2018,Goerg2019,Schweizer2019,Mil2020,Wang2021,Mildenberger2022,farrell2023scalable,angelides2023firstorder}. Such quantum simulators naturally incorporate many-body entanglement by working directly with the wave function, thereby reducing computational complexity from exponential to polynomial in the available resources due to quantum advantage. As such, large-scale robust and stabilized quantum simulators of high-energy physics hold the promise to probe nonperturbative far-from-equilibrium collider-relevant physics from first principles, providing temporal snapshots of their microscopic workings \cite{Bauer_review,dimeglio_review}. Furthermore, high-energy physics is an ideal field for driving progress in quantum simulators, as it offers a myriad of processes, such as neutrino (astro)physics and hadronization, where quantum advantage can prove crucial \cite{Berges_review}. This gives rise to a two-way street between high-energy physics and quantum simulation, where the former provides true tests of quantum advantage for the latter, and the latter delivers tangible devices to probe the former.

In recent years, cold-atom quantum simulators with optical superlattices have made a leap forward towards the large-scale quantum simulation of a model of $1+1$D lattice quantum electrodynamics (QED) \cite{Yang2020,Zhou2022,Su2022,Wang2022,Zhang2023}, facilitated by controlled schemes for the stabilization of gauge invariance against errors \cite{Halimeh2020a,Halimeh2020e,vandamme2021reliability,Halimeh_ShortReview}. Remarkably, QED in one spatial dimension serves as a prototype for three-dimensional QCD as they share many intriguing phenomena, from confinement to spontaneous pair production and string breaking. The scattering of excitations in $1+1$D models has attracted much attention in recent years, with numerical studies performed in quantum field theories \cite{Pichler2016, Rigobello2021, Chai2023, Belyansky2023} as well as in quantum spin models \cite{Vovrosh2022, Milsted2022}. However, a realistic protocol to realize such collision processes in the quantum simulator remains elusive. This is particularly concerning given that it is of strong interest to the community to advance state-of-the-art quantum simulators to the level where they can probe processes mimicking those in particle collision experiments \cite{dimeglio_review}, as this will bring these quantum simulators closer to their end goal of becoming complementary venues to particle colliders.

In this work, we propose particle collision experiments in a state-of-the-art optical-superlattice quantum simulator of a $\mathrm{U}(1)$ gauge theory. We introduce an experimentally feasible scheme to impart momenta on elementary and composite particles through holding potential walls, and then propose various collision experiments where rich physics can be probed; see Fig.~\ref{fig:kick}(a). Using numerical methods based on matrix product state (MPS) techniques~\cite{Uli_review, Paeckel_review, tenpy, mptoolkit}, we show that collisions of a wide range of energy scales can be accessed in our quantum simulator, and they give rise to numerous interesting phenomena, from string dynamics in- and out-of-equilibrium, to entropy production, and to the dynamical formation and breaking down of mesons.

\section{Lattice QED in a cold-atom quantum simulator}
We consider the canonical $1+1$D lattice QED Hamiltonian \cite{Kogut1975}
\begin{align}\nonumber
\hat{H}_\text{QED} 
&= - \frac{\tilde{\kappa}}{2a} \sum_{\ell=1}^{L-1}  \big(\hat{\psi}_{\ell}^\dagger \hat{U}_{\ell,\ell+1} \hat{\psi}_{\ell+1} + \text{H.c.}\big) \\\label{eq:qed}
& + m \sum_{\ell=1}^{L} (-1)^\ell \hat{\psi}_\ell^\dagger \hat{\psi}_\ell  +  \frac{a}{2} \sum_{\ell=1}^{L-1} \big(\hat{E}_{\ell,\ell+1}+\hat{E}_\text{bg}\big)^2,
\end{align}
employing the ``staggered fermion'' representation \cite{susskind1977} where opposite charges (particles and antiparticles) are placed on alternating sites, represented by the fermionic field operators $\hat{\psi}_{\ell},\hat{\psi}_{\ell}^\dagger$ on site $\ell$ of a chain with a total of $L$ sites. 
The first term of Hamiltonian~\eqref{eq:qed} is the kinetic energy of fermionic hopping coupled by the dynamical gauge field $\hat{U}_{\ell,\ell+1}$ on the link between sites $\ell$ and $\ell+1$, controlled by the lattice spacing $a$ with strength $\tilde{\kappa}$, and the second term is the fermionic occupation with rest mass $m$. Together, these two terms control the strength of the Schwinger pair production and annihilation process. The last term is the energy of gauge field coupling, where $\hat{E}_{\ell,\ell+1}$ is the electric field on the link between sites $\ell$ and $\ell+1$, and where we have introduced an additional homogeneous background electric field $\hat{E}_\text{bg}=g\theta/(2\pi)$, where $g$ is the gauge coupling strength and $\theta \in [0,2\pi)$ is the topological angle. This term tunes a confinement-deconfinement transition \cite{Coleman1975}.

In order to facilitate the numerical simulation and experimental implementation of QED on modern quantum simulators and, a so-called quantum link model (QLM) formulation is adopted \cite{Chandrasekharan1997,Wiese_review,Kasper2017}, where the dynamical gauge field and electric flux operators $\hat{U}_{\ell,\ell+1}$ and $\hat{E}_{\ell,\ell+1}$, respectively, are represented by spin operators: $\hat{U}_{\ell,\ell+1} \rightarrow \hat{S}_{\ell,\ell+1}^+/\sqrt{S(S+1)},\,\hat{E}_{\ell,\ell+1} \rightarrow g\hat{S}_{\ell,\ell+1}^z$. This representation satisfies the canonical commutation relation $[\hat{E}_{\ell,\ell+1},\hat{U}_{k,k+1}]=g\delta_{\ell k}\hat{U}_{k,k+1}$, and, in the Kogut--Susskind limit $S\to\infty$, the canonical commutation relation $[\hat{U}_{\ell,\ell+1},\hat{U}_{k,k+1}^\dagger]=0$. We further perform the particle-hole transformation \cite{Yang2016} for odd sites $\ell$: $\hat{\psi}_{\ell} \rightarrow \hat{\psi}_{\ell}^\dagger$ and $\hat{S}_{\ell-1,\ell}^z \rightarrow -\hat{S}_{\ell-1,\ell}^z$, rendering Eq.~\eqref{eq:qed} as
\begin{align}\nonumber
\hat{H}_\text{QLM} 
= & -\frac{\kappa\sqrt{3}}{4a\sqrt{S(S+1)}} \sum_{\ell=1}^{L-1}  \big(\hat{\psi}_{\ell} \hat{S}_{\ell,\ell+1}^{+} \hat{\psi}_{\ell+1} + \text{H.c.}\big) 
\\\nonumber
& + m\sum_{\ell=1}^{L}\hat{\psi}_\ell^\dagger \hat{\psi}_\ell   + \frac{a g^2}{2} \sum_{\ell=1}^{L-1}  \big(\hat{S}_{\ell,\ell+1}^z\big)^2 \\\label{eq:qlm}
&- a \chi\sum_{\ell=1}^{L-1}  (-1)^\ell \hat{S}_{\ell,\ell+1}^z, 
\end{align}
where $\kappa=2\tilde{\kappa}/\sqrt{3}$ is used to obtain a coupling constant of $\kappa/2$ in the case of $S=1/2$, which will become convenient when we restrict to $S=1/2$ later on.
The continuum limit of QED is recovered at $S \rightarrow \infty$ and $a \rightarrow 0$ \cite{Buyens2017, Banuls2020, Zache2021achieving,Halimeh2021achieving}. The last term is a staggering potential on the gauge fields, which realizes the background electric field $\hat{E}_\text{bg}$ that tunes the topological $\theta$-angle, with $\chi=g^2(\theta-\pi)/(2\pi)$ \cite{Surace2020, Halimeh2022tuning, Cheng2022tunable}. 

The $\mathrm{U}(1)$ gauge transformation is generated by the local Gauss's law operators
\begin{align}\label{eq:gauss}
\hat{G}_\ell = (-1)^\ell \left[ \hat{S}_{\ell,\ell+1}^z + \hat{S}_{\ell-1,\ell}^z + \frac{\hat{\psi}_\ell^\dagger \hat{\psi}_\ell+(-1)^\ell}{2} \right],
\end{align}
which commute with the QLM Hamiltonian~\eqref{eq:qlm}: $[\hat{G}_\ell, \hat{H}_\text{QLM}]=0,\,\forall \ell$, underlining local gauge invariance in that particle hopping must be accompanied by concomitant changes in the local electric fields such as to preserve Gauss's law. As per convention, we choose to work in the \textit{physical} gauge sector of states $\ket{\Psi}$ satisfying $\hat{G}_\ell \ket{\Psi}=0,\, \forall \ell$. 

Recently, the spin-$1/2$ $\mathrm{U}(1)$ QLM has been experimentally realized in a large-scale Bose--Hubbard quantum simulator \cite{Yang2020,Zhou2022,Su2022,Wang2022,Zhang2023}, and we shall henceforth restrict to this case of $S=1/2$. The local electric field spans the basis $\{ \ket{\triangleleft},\ket{\triangleright} \}$ encoding two eigenstates of the spin-$1/2$ operator $\hat{S}_{\ell,\ell+1}^z$, see Fig.~\ref{fig:kick}(b). The gauge coupling term $\propto (\hat{S}_{\ell,\ell+1}^z)^2=\mathds{1}_{\ell,\ell+1}/4$ becomes an inconsequential constant energy term that does not contribute to the dynamics at $S=1/2$. The spin-$1/2$ $\mathrm{U}(1)$ QLM is deconfined at $\theta=\pi$ and hosts Coleman's phase transition at the critical mass $m_\text{c}=0.3275\kappa$ \cite{Coleman1976}, which is related to the spontaneous breaking of a global $\mathbb{Z}_2$ symmetry connected to charge conjugation and parity symmetry conservation; see Fig.~\ref{fig:kick}(c). For $m \rightarrow + \infty$, the ground state manifold of $\hat{H}_\text{QLM}$ corresponds to the two degenerate vacua $\ket{\ldots \triangleleft,\varnothing,\triangleleft,\varnothing,\triangleleft \ldots}$ and $\ket{\ldots \triangleright,\varnothing,\triangleright,\varnothing,\triangleright \ldots}$, where $\varnothing$ represents the absence of matter at a site. In this case ($\theta=\pi$), no string tension is present between a particle-antiparticle pair. Tuning the $\theta$-angle away from $\pi$ creates an additional background electric field $\hat{E}_\text{bg}$ that explicitly breaks this global $\mathbb{Z}_2$ symmetry, creating an energy difference between the two electric fluxes $\{ \ket{\triangleleft},\ket{\triangleright} \}$. As a result, a particle-antiparticle pair connected by a string of $D$ electric fluxes $\ket{\triangleleft}$ experiences the string energy $\chi D$ that increases linearly with $D$. Subsequently, the spin-$1/2$ QLM becomes a confining theory. A particle-antiparticle pair in the confined $1+1$D QED theory forms a two-particle bound state, analogous to a meson in $3+1$D QCD, which is a composite particle made of a quark-antiquark pair formed by the gluon flux tube connecting them \cite{Coleman1975, Wilson1974a}.

The quantum simulator used to experimentally realize the spin-$1/2$ QLM is governed by the Bose--Hubbard  model (BHM) Hamiltonian \cite{Yang2020}
\begin{align}\nonumber
\hat{H}_\text{BHM} =&-J\sum_{j=1}^{N-1} \left( \hat{b}^\dagger_{j}\hat{b}_{j+1}+ \hat{b}^\dagger_{j+1}\hat{b}_{j}\right)  \\\label{eq:bhm}
&+\frac{U}{2}\sum_{j=1}^N \hat{n}_j\left(\hat{n}_j-1\right)+\sum_{j=1}^N \mu_j \hat{n}_j,
\end{align}
with $\hat{b}_{j}$, $\hat{b}^\dagger_{j}$ the bosonic field operators, $J$ the tunneling strength between neighboring sites, $U$ the on-site interaction, and $N=2L$ the total number of sites in the quantum simulator. The chemical potential term is used to engineer the correlated hopping process that implements the gauge theory Hamiltonian~\eqref{eq:qlm} at $S=1/2$. It takes the form $\mu_j = j  \Delta + (-1)^j \delta/2 + \text{sin}(j \pi/2)\chi/2$, where $\Delta$ is a linear tilt used to suppress long-range single atom tunneling \cite{Halimeh2020d}, $\delta$ is a staggering potential generated by a period-$2$ optical superlattice separating the system into two sublattices, the ``matter'' sites are denoted as $j_\text{M} \rightarrow \ell$ ($j$ even) and the ``gauge'' sites are denoted as $j_\text{G} \rightarrow (\ell,\ell+1)$ ($j$ odd). For $\delta \sim U/2 \gg J$, we identify the resonant second-order correlated hopping process $-\frac{\kappa}{2\sqrt{2}}[\hat{b}_{j-1}(\hat{b}^\dagger_{j})^2\hat{b}_{j-1}+ \text{H.c.}]$ (with $j$ odd) where single bosons on neighboring matter sites annihilate (create) to form a doublon (hole) on the gauge site in between; see Fig.~\ref{fig:kick}(a). As a result, the effective Hamiltonian takes the form of Eq.~\eqref{eq:qlm}, and we identify $\kappa \approx 8\sqrt{2}J^2/U$ and $m=\delta/2-U$ by using second-order perturbation theory \cite{Yang2020}. The confining term $\propto \chi$ is a staggered potential on gauge sites generated by a period-$4$ optical superlattice \cite{Halimeh2022tuning}, which breaks the degeneracy between the two vacua $\ket{\ldots \triangleleft,\varnothing,\triangleleft,\varnothing,\triangleleft,\varnothing,\triangleleft \ldots} \leftrightarrow \ket{\ldots 2,0,0,0,2,0,0 \ldots}$ and $\ket{\ldots \triangleright,\varnothing,\triangleright,\varnothing,\triangleright,\varnothing,\triangleright \ldots} \leftrightarrow \ket{\ldots 0,0,2,0,0,0,2 \ldots}$, where here we show their bosonic representation on the optical superlattice.

For all numerical simulations we have performed in this paper, we use the experimentally tested parameters $J=58$~Hz, $U=1368$~Hz, $\Delta=57$~Hz \cite{Zhou2022}, and subsequently $\kappa \approx 28$~Hz.

\begin{figure}[htb!]
	\centering
	\includegraphics[width=\linewidth]{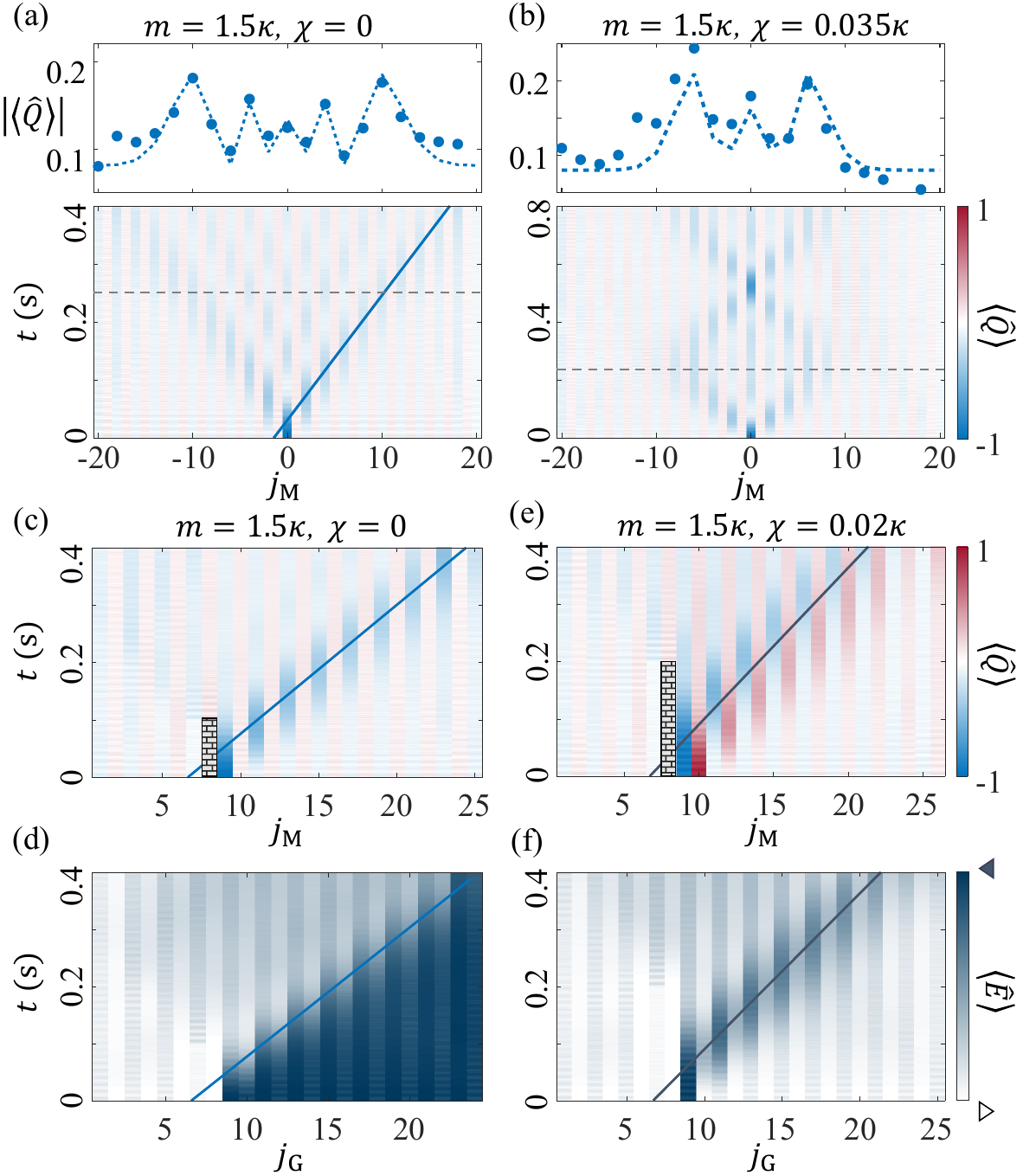}
	\caption{(Color online).
        \textbf{Single particle quantum walk and momentum initialization}
        (a) Single particle quantum walk on matter sites. The upper panel shows the wavefront at $t=0.25$s (gray dashed line in lower panel), and fitted to the Bessel function Eq.~\eqref{eq:qw} (blue dashed line) where $\Tilde{t}_\text{fit}=2.1(1)$~Hz, close to $\Tilde{t}=2.33$~Hz predicted by the second order perturbation theory. The propagation speed of the wavefront is linearly fitted to be $46(2) ~\text{sites}/\text{s}$. 
        (b) Bloch oscillation of a single particle. The wavefront at $t=0.25$s (gray dashed line in lower panel) is fitted to Eq.~\eqref{eq:bo} (blue dashed line) with $\chi_\text{fit}=0.96(2)\,\text{Hz} \approx 0.035\kappa$.
        (c) and (d): The moving wave packet of a single particle created by reflection on the barrier and propagating to the right. The group velocity initiated is extracted by a linear fit, where we find $v_\text{g}^{(i)}=47.9(8)~\text{sites}/\text{s}$ (blue line). The electric flux $\langle \hat{E} \rangle$ changes accordingly with Gauss's law as the particle propagates.
        (e) and (g): A moving meson wave packet. The group velocity is fitted to be $37.4(4)~\text{sites}/\text{s}$ (dark blue line). 
        }
	\label{fig:MovingWavePacket}
\end{figure}

\section{Cold-atom ``particle accelerator''}
\label{sec:ParticleAccelerator}

The basic ingredients of particle collision experiments are the spatially localized moving wave packets of elementary or composite particles \cite{Weinberg_book}. An elementary particle or antiparticle excitation in the vacuum background can be expressed as $\ket{\dots \triangleright,\varnothing,\triangleright,-,\triangleleft,\varnothing,\triangleleft \dots}$ or $\ket{\dots \triangleleft,\varnothing,\triangleleft,+,\triangleright,\varnothing,\triangleright \dots}$, which corresponds to the state $\ket{\dots 2,0,0,\textit{1},0,0,2 \dots}$ in the Bose--Hubbard model (a single boson $\ket{\textit{1}}$ on an odd matter site for a particle and an even matter site for an antiparticle), see also Fig.~\ref{fig:kick}(b). A particle-antiparticle meson excitation $\ket{\dots \triangleright,\varnothing,\triangleright,-,\triangleleft,+,\triangleright,\varnothing,\triangleright \dots}$ corresponds to the state $\ket{\dots 2,0,0,\textit{1,0,1},0,0,2 \dots}$. 

We first consider the regime $m \gg m_\text{c}$ where the rest mass dominates, and spontaneous Schwinger pair production is exponentially suppressed. The (anti)particle hopping is a second-order process in the QLM with a virtual pair creation as intermediate step,
see Fig.~\ref{fig:kick}(a) and Appendix~\ref{app:PT}. 
Therefore, the low-energy effective Hamiltonian of the quantum link model \eqref{eq:qlm} becomes
\begin{align}\nonumber
\hat{H}_\text{(A)P} =  
&-\Tilde{t} \sum_{\ell_\text{(A)P}} \big(\hat{\psi}_{\ell_\text{(A)P}}^\dagger \hat{\psi}_{\ell_\text{(A)P}+1} + \text{H.c.}\big) \\
\label{eq:eff}
&+(-1)^{\sigma_{\text{(A)P}}} 2\chi \sum_{\ell_\text{(A)P}} \ell_\text{(A)P} \hat{\psi}_{\ell_\text{(A)P}}^\dagger \hat{\psi}_{\ell_\text{(A)P}},
\end{align}
where $\hat{\psi}_{\ell_\text{(A)P}}, \hat{\psi}_{\ell_\text{(A)P}}^\dagger$ are fermionic fields on (anti)particle sublattice sites ($\ell_\text{P}$ for $\ell$ odd, $\ell_\text{A}$ for $\ell$ even). The background electric field created by the confining potential $\chi$ can be understood as an effective linear tilt, with $\sigma_{\text{A}}=0$ and $\sigma_{\text{P}}=1$ which makes the tilt positive for the antiparticle and negative for the particle.
We identify this effective (anti)particle tunneling strength to be $\Tilde{t} = \kappa^2/(8ma^2)$ by using second-order perturbation theory (Appendix~\ref{app:PT}).

\subsection{Single particle quantum walk}
Before we create moving wave packets, we will first look at the dynamics of a single particle localized on a lattice site, which is a coherent superposition of all momentum eigenstates within the first Brillouin zone. The localized wave packet has an equal probability of tunneling in both directions. At $\chi=0$, a single (anti)particle undergoes a quantum walk, analogous to a free electron in a homogeneous lattice \cite{Preiss2015}.  The result is a light-cone-shaped transport, and the wave packet delocalizes, as shown in Fig.~\ref{fig:MovingWavePacket}(a).

To characterize the quantum walk, we numerically calculate the expectation value of the charge density operator in the Bose--Hubbard quantum simulator,
\begin{align}
    \langle \hat{Q}_{j_\text{M}}(t) \rangle = \langle \hat{Q}_\ell(t) \rangle=\bra{\Psi(t)}\hat{Q}_\ell\ket{\Psi(t)},
\end{align}
with $\ket{\Psi(t)}=\text{exp}(-i \hat{H}_\text{BHM} t)\ket{\Psi_0}$ and $\hat{Q}_\ell=(-1)^\ell \hat{\psi}_{\ell}^\dagger \hat{\psi}_{\ell}$, where $\ket{\Psi_0}$ is the initial state and $t$ is evolution time. We choose $m=1.5\kappa$ for the numerical simulations as it is large enough to suppress spontaneous pair creation and maintain the mapping to the effective Hamiltonian \eqref{eq:eff} while keeping the dynamics fast enough for experimental implementations with limited coherence time.

For a single particle, the charge density wavefront on the particle sublattice can be characterized by the Bessel function of the first kind $\mathcal{J}_{\ell_\text{P}}$ \cite{Hartmann2004}
\begin{equation}
\label{eq:qw}
    \big\lvert\langle \hat{Q} (t) \rangle\big\rvert_{\ell_\text{P}} = A_0\big\lvert\mathcal{J}_{\ell_\text{P}}(2 \times 2\pi \Tilde{t} \times t)\big\rvert^2+B_0.
\end{equation}
The first-order dynamics of spontaneous pair production and annihilation in the QLM, as well as the direct tunneling in the BHM, result in a shift in the background and the reduction of amplitude, which we account for by adding two extra parameters $A_0$ and $B_0$. The effective tunneling is fitted to be $\Tilde{t}_\text{fit}=2.1(1)$~Hz; see the dashed curve in Fig.~\ref{fig:MovingWavePacket}(a). The fitting result is slightly smaller than $\Tilde{t}=2.33$~Hz predicted by the approximate second-order perturbation theory, as the actual dispersion relation deviates slightly from the sinusoidal function expected from the effective model in Eq.~\eqref{eq:eff}. This is mainly due to the first-order pair-creation dynamics in the QLM, and thus the actual dispersion is slightly different from the sinusoidal dispersion expected from Eq.~\eqref{eq:eff} (see also Fig.~\ref{fig:single-particle-spectra}(a)). The propagation speed of the outer wavefront is fitted to be $46(2) \text{sites}/\text{s}$, while the maximum group velocity estimated from the ground band of the effective model is $v_\text{g}^\text{max}=58.6~\text{sites}/\text{s}$, in particular Eq.~\eqref{eq:vg}. In fact, the group velocity fit $46(2)~ \text{sites}/\text{s}$ actually matches pretty closely to the maximum group velocity from the MPS calculation in Fig.~\ref{fig:single-particle-spectra}(b) $v_\text{g}^\text{MPS}=48.3~\text{sites}/\text{s}$; see Appendix~\ref{app:PT}.

For $\chi \neq 0$, although there is an external force acting on the (anti)particle, a net transport in the lattice is not possible, as the maximum kinetic energy of the particle is limited by the bandwidth. As a result, the (anti)particle undergoes Bloch oscillations, as shown in Fig.~\ref{fig:MovingWavePacket}(b). In this case, the charge density can be characterized by modifying the argument of the Bessel function as \cite{Hartmann2004}
\begin{align}
\label{eq:bo}
    \big\lvert\langle \hat{Q} (t)\rangle\big\rvert_{\ell_\text{P}}= A_0\bigg\lvert\mathcal{J}_{\ell_\text{P}}\bigg[\frac{4\Tilde{t}}{2\chi} \text{sin}(2\chi\pi t)\bigg]\bigg\rvert^2+B_0.
\end{align}
The dashed line in Fig.~\ref{fig:MovingWavePacket}(b) is a fit to Eq.~\eqref{eq:bo}, where we find $\chi_\text{fit}=0.96(2)\,\text{Hz}$, which agrees well with $\chi=0.035\kappa=0.98\,\text{Hz}$. We notice an imbalance of the wavefront related to the sign of $\chi$, which we attribute to higher-order processes that are not captured by the effective model.

\begin{figure}
	\centering
	\includegraphics[width=\linewidth]{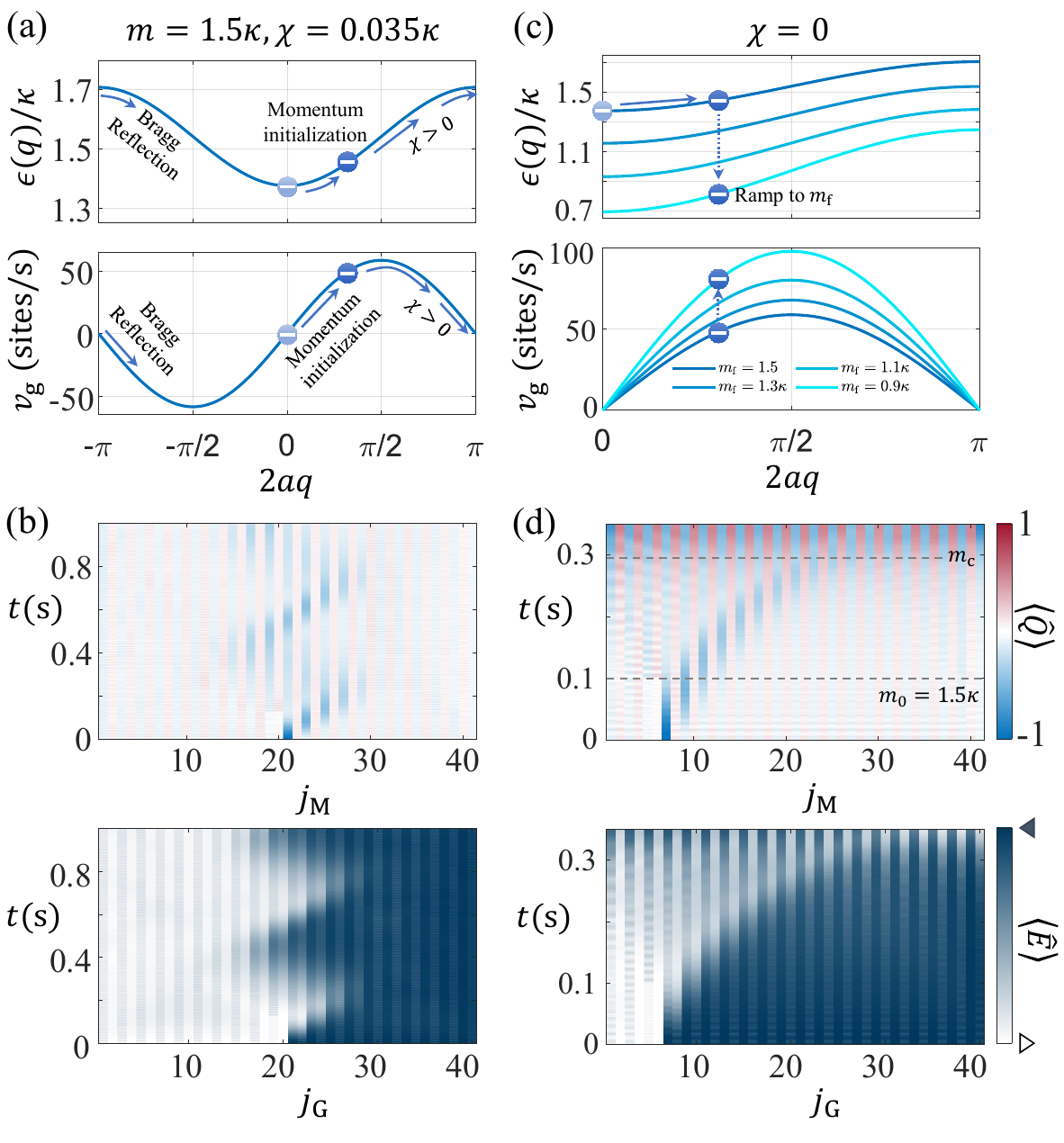}
	\caption{(Color online).
        \textbf{Real space Bloch oscillation and particle acceleration.}
        (a) Upper panel: Ground band dispersion relation of a single particle (Eq.~\eqref{eq:eff-dispersion}). Lower panel: Group velocity calculated from the dispersion (Eq.~\eqref{eq:vg}). Although the quasi-momentum increases linearly due to the external force induced by $\chi$, the maximum group velocity is bounded by the bandwidth, and the wave packet undergoes Bloch oscillation. 
        (b) Real-space Bloch oscillation of a single-particle moving wave packet. 
        (c) Ground band of a single particle with different mass (Eq.~\eqref{eq:eff-dispersion}). A particle with a smaller mass has lower rest energy and higher velocity for the same quasi-momentum.
        (d) Particle acceleration by tuning the rest mass. We prepare the moving wave packet from 0 to $0.1$~s at $m=1.5\kappa$, and starting at $0.1$~s, we linearly ramp down the mass to $m=0$ in $0.25$~s. Before reaching the critical point $m_\text{c}=0.3275\kappa$, we find the particle undergoes continuous acceleration. Below the critical point, we can no longer identify the single particle wave packet due to spontaneous pair creation in the vacuum background.
        }
	\label{fig:BlochOscillation}
\end{figure}

\subsection{Momentum initialization}

To explore collision dynamics, previous theoretical works have typically generated uni-directional moving wave packets by numerically building a superposition of momentum eigenstates \cite{Rigobello2021, Vovrosh2022, Milsted2022}. Although this method is straightforward in numerical simulations, it is quite challenging in a cold-atom experiment. Here, we demonstrate a simple scheme to prepare such moving wave packets in experiment with a potential barrier, as illustrated schematically in Fig.~\ref{fig:kick}(a). The potential barrier can be achieved by single-site addressing with a blue-detuned light potential through the high-resolution objective in a quantum gas microscope experiment \cite{Weitenberg:2011, Islam2015, Zhang2023}. An (anti)particle excitation localized in space is a coherent superposition of momentum eigenstates with momentum components centered at $q=0$. The barrier placed left (right) to the original wave packet reflects the left (right)-moving momentum components to the right (left), thus shifting the center of the momentum superposition to a finite value, effectively creating a moving wave packet. We show that this method can work for elementary particles in Fig.~\ref{fig:MovingWavePacket}(c,d) as well as composite particles (mesons) in Fig.~\ref{fig:MovingWavePacket}(e,f). 

We note that, unlike a free-electron theory, here the particle hopping is coupled by the gauge fields, with corresponding flips of electric fluxes as the particle moves, such that the Gauss's law Eq.~\eqref{eq:gauss} is always satisfied, see Fig.~\ref{fig:MovingWavePacket}(d). In our numerical simulations, the barrier is encoded as a local chemical potential placed on the two sites left to the particle with $\mu_\text{address}=10\kappa \approx 120\Tilde{t}$ large enough to suppress the effective hopping. We remove the barrier at $0.1$~s, after the wave packet has moved away. We perform a linear fit of the particle trajectory and find the initial group velocity to be $v_\text{P}^\text{(i)}=47.9(8)~\text{sites}/\text{s}$, close to the speed of wavefront in the quantum walk without the barrier, shown in Fig.~\ref{fig:MovingWavePacket}(a).

For a meson state, pair hopping is a fourth-order process, illustrated on the right of Fig.~\ref{fig:kick}(a). The barrier forbids the antiparticle from hopping to the right, meanwhile the particle hops to the left, and the antiparticle follows afterward. In the presence of the confining potential, the external force exerted by the background electric field on the particle is $F_\text{P} \propto \chi$ pointing towards the right, and for the antiparticle, it is $F_\text{A} \propto -\chi$ pointing toward the left. Therefore, no net force is exerted on the center of mass of the pair. However, the same background field creates a string energy proportional to the inter-particle distance $D$. Subsequently, the hopping of the particle increases the string energy, making it energetically favorable for the antiparticle to follow. As a result, the particle-antiparticle pair moves together as a composite particle. We show the case of $m=1.5\kappa$ and $\chi=0.02\kappa$ in Fig.~\ref{fig:MovingWavePacket}(e,f). We perform a linear fit and find the group velocity of the meson to be $v_\text{g}^\text{M}=37.4(4)~\text{sites}/\text{s}$. To benchmark the speed of the meson, we calculate the meson band structure using the MPS excitation ansatz (Fig.~\ref{fig:meson-spectra}; see Appendix~\ref{sec:meson-excitations} for details). From the ground band dispersion, we find that the maximum group velocity of a meson is about $40~\text{sites}/\text{s}$, which agrees well with the extracted group velocity.

\begin{figure}
	\centering
	\includegraphics[width=\linewidth]{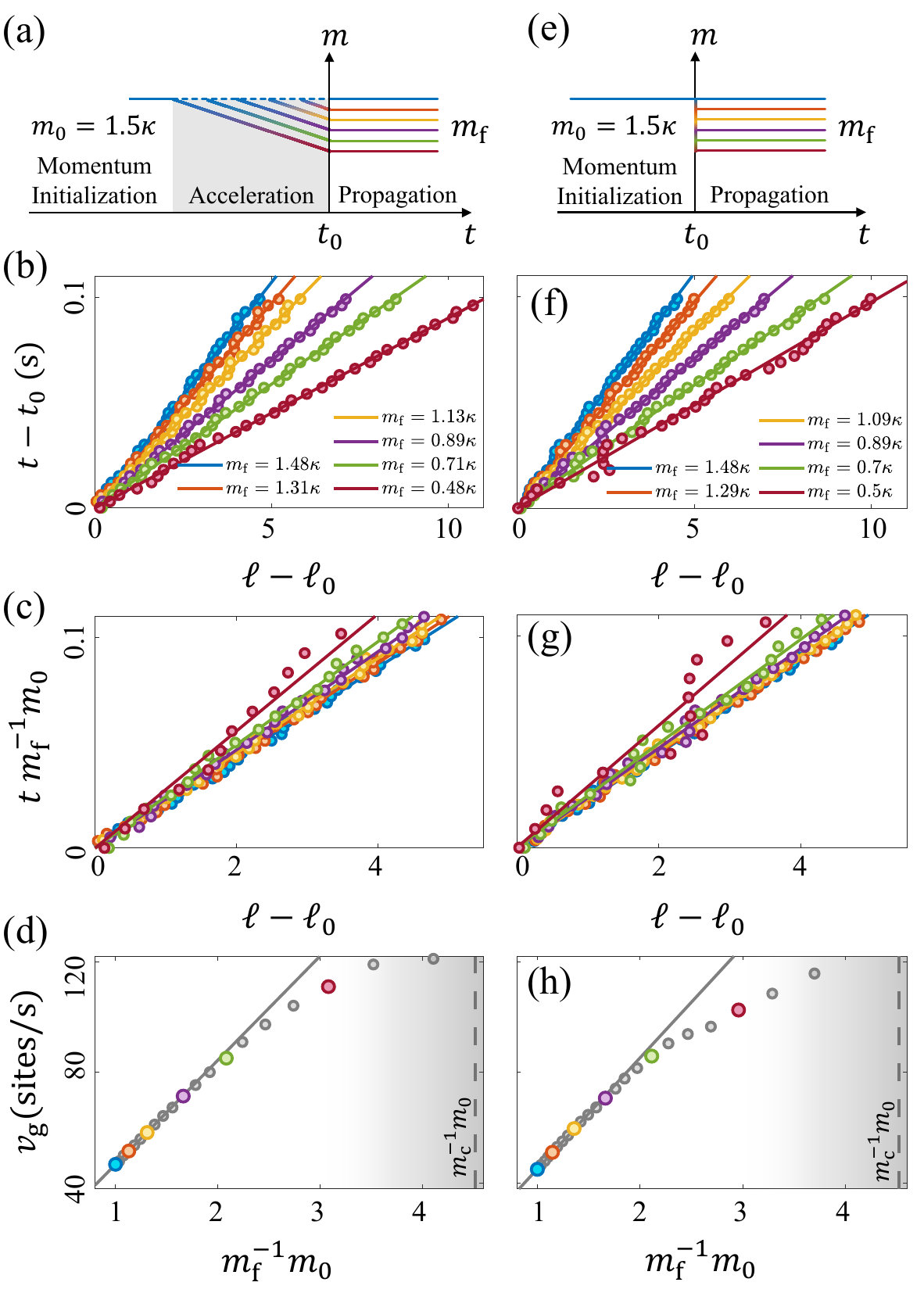}
	\caption{(Color online).
        \textbf{Analysis of the particle acceleration.}
        (a) Particle acceleration by ramping mass. After preparing the moving wave packet at $m_0=1.5\kappa$ in $0.1$~s, we linearly ramp down the mass to $m_\text{f}$ and let the wave packet propagate at the final mass. 
        (b) We extract the position of the wave packet at different times by a Gaussian fit, and then do a linear fit for each curve to extract the group velocity. 
        (c) By rescaling the propagation time with $m_\text{f}$, the space-time trajectory of wave packets with different final masses collapse onto a single line.
        (d) Group velocity is inversely proportional to the final mass $v_\text{g} \propto m_\text{f}^{-1}$ for $m_\text{f} \gg m_\text{c}$, and deviates from the linear relationship when approaching the critical mass $m_\text{c}$. 
        (e) Particle acceleration by quenching the mass from $m_0=1.5\kappa$ to $m_\text{f}$ at $t=0$. 
        (f)-(h): Same as (b)-(d). With the quench protocol, the group velocity still follows the linear relationship for larger $m_\text{f}$. However, when approaching $m_\text{c}$, the deviation is faster than the ramping protocol, as the quench creates more particle excitations in the vacuum background.
        }
	\label{fig:ParticleAccelerator}
\end{figure}

\subsection{Particle acceleration} 
The most essential feature of a real particle collider is the ability to accelerate the particles close to the speed of light, so that the collision creates a far-from-equilibrium system with such a high energy density that spontaneous particle production becomes possible.
To explore the relevant physics in this high energy scale, we investigate particle acceleration in our quantum simulator.

Naively, one would expect the background electric field created by potential $\chi$ to accelerate the (anti)particles. While this would be achievable in the continuum limit, on a lattice, the particles eventually undergo Bloch oscillations \cite{Milsted2022}; see Fig.~\ref{fig:BlochOscillation}(a) and (b).

However, utilizing the tunability of the quantum simulator, particle acceleration can be achieved by tuning down the rest mass. For $m \gg m_\text{c}$, the effective tunneling is inversely proportional to the mass $m$, $\Tilde{t} \propto m^{-1}$, and subsequently the group velocity $v_\text{g}\propto m^{-1}$ (the quasi-momentum of the wave packet remains constant during time evolution for $\chi=0$), see Fig.~\ref{fig:BlochOscillation}(c). Moreover, the tunable energy scale between the rest mass and the kinetic energy makes it possible to access regimes where spontaneous pair production dominates the dynamics. 

After momentum initialization at a constant mass $m_0=1.5\kappa$ from $0$ to $0.1$~s, we ramp down the mass from $m_0$ to $0$ in $0.25$~s, and observe a continuous acceleration of the particle before reaching the critical mass $m_\text{c}$ of Coleman's phase transition, see Fig.~\ref{fig:BlochOscillation}(d). As the critical point is approached, we find particle-antiparticle pair production in the vacuum background dominates and the initial wave packet is no longer observable.

To benchmark the acceleration, after ramping down the mass to a final mass $m_\text{f}$, we let the wave packet propagate at $m_\text{f}$ starting at $t_0$, see Fig.~\ref{fig:ParticleAccelerator}(a). After this point, the wave packet propagates at a constant velocity. We extract the position of the wave packet every $0.003$s by Gaussian fits and then perform a linear fit to find the group velocity, see Fig.~\ref{fig:ParticleAccelerator}(b). As expected from the band structure, the group velocity $v_\text{g}$ rises linearly with $m_\text{f}^{-1}$ for $m_\text{f} \gtrsim \kappa$, and deviates from the linear relationship approaching the critical mass $m_\text{c}$, see Fig.~\ref{fig:ParticleAccelerator}(d). By rescaling the evolution time of each curve with its final mass $m_\text{f}^{-1}$, we also find all the trajectories collapse onto a single line, while small deviations can be found for $m_\text{f} \sim m_\text{c}$, see Fig.~\ref{fig:ParticleAccelerator}(c). 

A more interesting protocol of acceleration is instantaneously quenching the mass to $m_\text{f}$ at $t=t_0$, which is a global quench that brings the system out of equilibrium \cite{Zhou2022}, see Fig.~\ref{fig:ParticleAccelerator}(e)-(h). For less violent quenches ($m_\text{f} \gtrsim \kappa$) we find the group velocity maintains $v_\text{g}\propto m^{-1}$, but deviates from the linear relationship faster than the ramping protocol when approaching $m_\text{c}$. This is expected, since the quench creates more particle-antiparticle excitations in the vacuum background than the ramp. In this case, the wave packets can no longer be extracted below $m_\text{f}=0.375\kappa$ ($m_\text{f}^{-1}m_0=4$ in Fig.~\ref{fig:ParticleAccelerator}(h)).

\begin{figure}
	\centering
	\includegraphics[width=\linewidth]{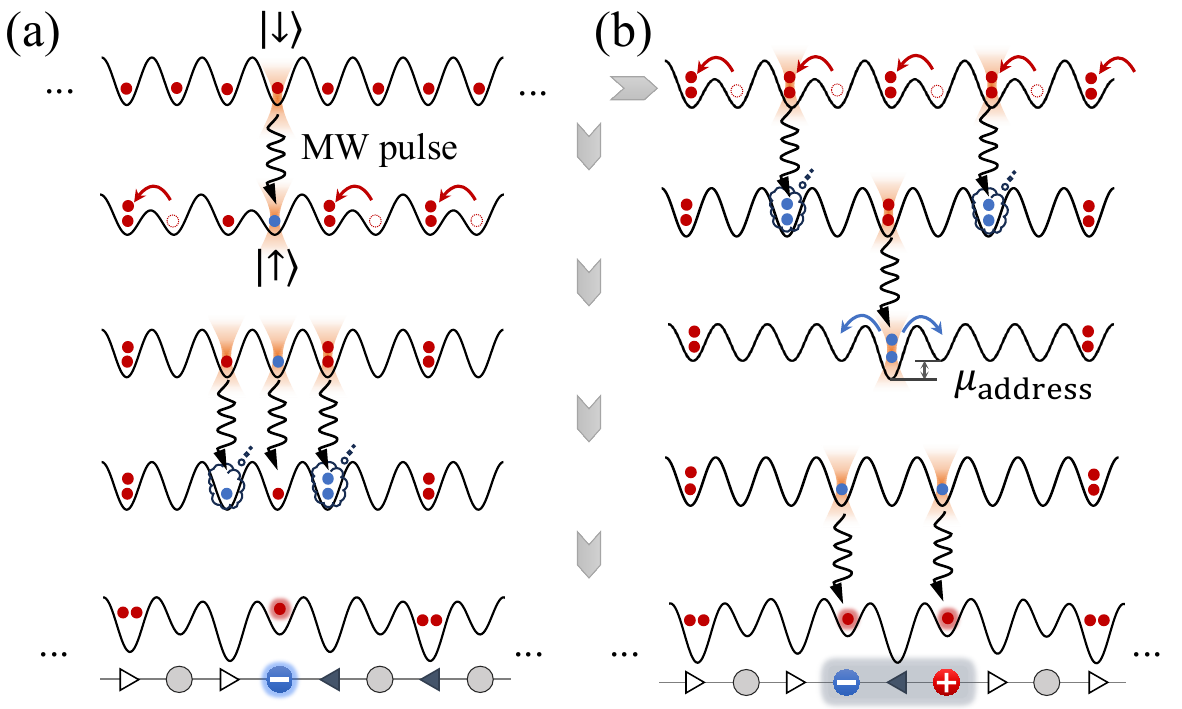}
	\caption{(Color online).
        \textbf{Initial state preparation and momentum initialization}
        Red dots denote atoms in hyperfine state $\ket{\downarrow}=\ket{F=1,m_F=-1}$, while blue dots denotes atoms in $\ket{\uparrow}=\ket{F=2,m_F=-2}$. Addressing tweezer beams induces local AC-stark shift on $\ket{\uparrow}$ state, such that the addressed atoms can be individually transferred between $\ket{\downarrow}$ and $\ket{\uparrow}$ with a microwave pulse. 
        (a) Preparation of a single (anti)particle initial state. 
        (b) Preparation of a particle-antiparticle pair.
        }
	\label{fig:init}
\end{figure}

\subsection{Initial state preparation}

Here we describe how these elementary particles and particle-antiparticle pairs can be prepared in a cold-atom quantum simulator. We consider the well-tested experiment with $^{87}\text{Rb}$ atoms in optical superlattices \cite{Yang2020}. The proposed experiment starts with a $\langle \hat{n} \rangle =1$ Mott insulator state, where all atoms are prepared in the hyperfine state $\ket{\downarrow}=\ket{F=1,m_F=-1}$, see Fig.~\ref{fig:init}(a). To prepare a single particle excitation, we first address a single atom with a $\sigma^-$ polarized optical tweezer at a wavelength of 787.55nm, which creates a light shift acting only on internal state $\ket{\uparrow}=\ket{F=2,m_F=-2}$ \cite{Weitenberg:2011, Zhang2023}. The addressed atoms can then be flipped to $\ket{\uparrow}$ with a resonant microwave field. While this atom is pinned by the addressing beam, we merge the remaining atoms into odd sites with a superlattice \cite{Yang2020a}, creating the state $\ket{\dots 2,0,1,1,2,0,2,0 \dots}$. We now project $\sigma^-$ tweezers onto the single atoms along with alternating doublons on every four sites, and flip them to $\ket{\uparrow}$ which are removed by a resonant laser. The resulting state is a single (anti)particle in the vacuum background $\ket{\dots 2,0,0,1,0,0,2,0 \dots}$.

To prepare a particle-antiparticle pair, we first create a $\mathbb{Z}_2$ ordered product state $\ket{\dots 2,0,2,0 \dots}$ from the $\langle \hat{n} \rangle =1$ Mott insulator with the superlattice, see Fig.~\ref{fig:init}(b). By addressing and removing alternating doublons with an array of $\sigma^-$ tweezers, we create a $\mathbb{Z}_4$ ordered state $\ket{\dots 2,0,0,0,2,0,0,0 \dots}$ corresponding to a vacuum state in the gauge theory. We now address a single doublon with the $\sigma^-$ tweezer beam and flip both atoms to $\ket{\uparrow}$ state, the local chemical potential $\mu_\text{address}$ created by the tweezer tunes the rest mass locally. By tuning the intensity of this addressing tweezer we can tune the local rest mass $m_\text{address}$ to $0$ and initiate the second-order correlated tunneling $\ket{0,2,0} \rightarrow \ket{1,0,1}$ to split this doublon into a pair of single atoms on neighboring matter sites, corresponding to a local particle-antiparticle pair in the gauge theory.

\section{Collision dynamics}

In this section, we demonstrate the rich physics that can be probed with particle collisions in the quantum simulator. 

\subsection{Particle--antiparticle collision}
\label{sec:epcollider}

\begin{figure}[htb!]
	\centering
	\includegraphics[width=\linewidth]{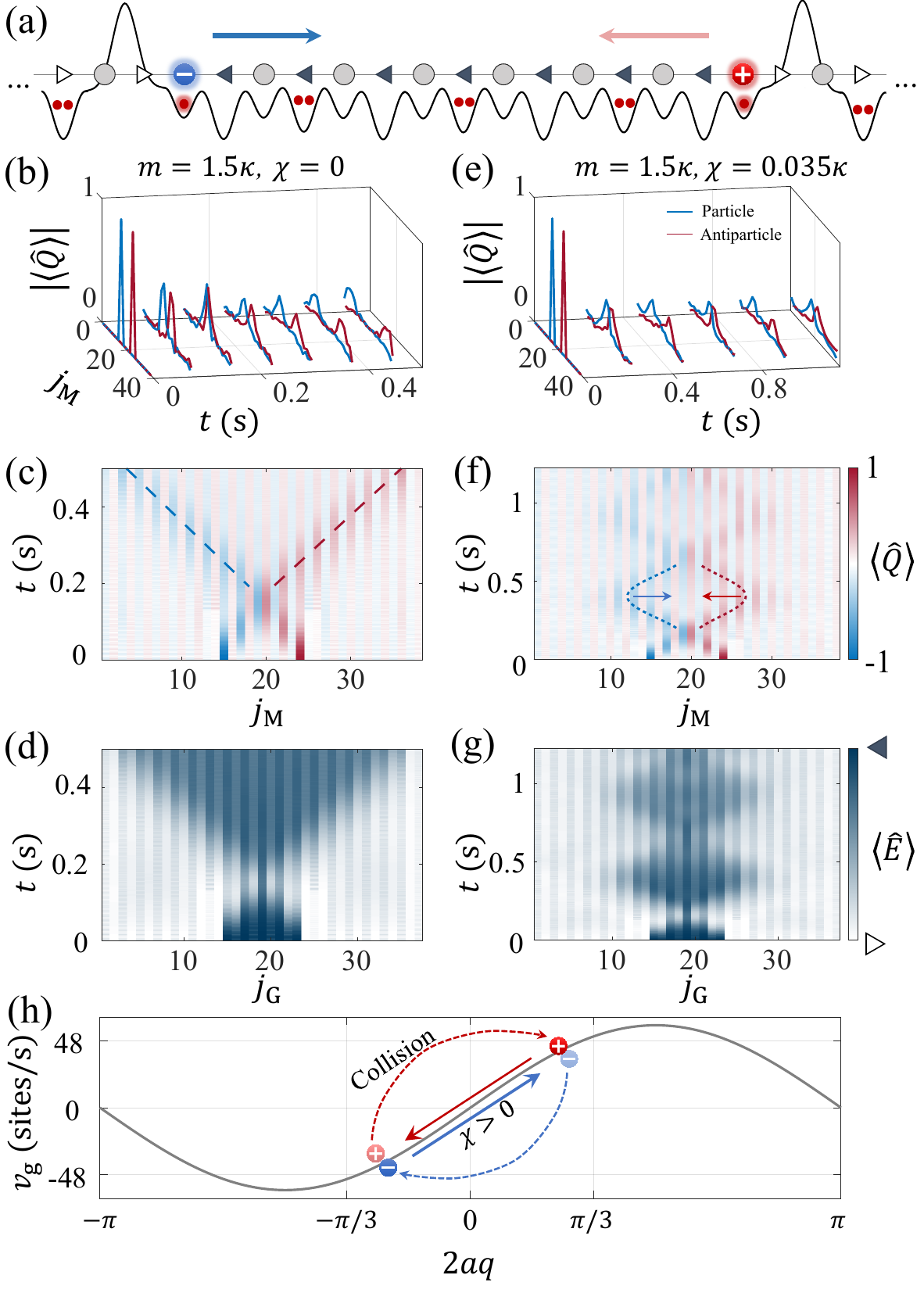}
	\caption{(Color online).
        \textbf{Particle-antiparticle collisions in the large mass limit.}
        (a) A schematic illustration of particle-antiparticle collision in the quantum simulator. 
        (b) and (e): Expectation value of charge density on matter sites $\langle \hat{Q} \rangle$ illustrating the collision of moving particle (blue) and antiparticle (red) wave packets. 
        (c) and (f): Same data as in the upper row with color plots. 
        (d) and (g): Expectation value of electric flux on gauge sites $\langle \hat{E} \rangle$, illustrating the string dynamics. 
        (b)-(d) Collision dynamics in the deconfined case ($\chi=0$). The group velocity of the particle and antiparticle is fitted to be $v_\text{P}^\text{(f)}=-47.4(5)~\text{sites}/\text{s}$ (dashed blue line), and $v_\text{A}^\text{(f)}=48(1)~\text{sites}/\text{s}$ (dashed red line), respectively. The particle and antiparticle undergo an elastic collision. 
        (e)-(g) Collision dynamics in the confined case ($\chi=0.035\kappa$). The confining potential results in higher energy for the $\ket{\triangleleft}$ electric flux, thus creating a string tension between the particle-antiparticle pair, leading the multiple collisions. 
        (h) Illustration of collision dynamics in the momentum space. After the collision, the particle and antiparticle move apart in opposite directions. When $\chi>0$, they experience a constant attractive force $\propto \chi$ which accelerates them towards each other. 
        }
	\label{fig:StringDynamics}
\end{figure}

We first consider the low-energy collision between a particle and an antiparticle, see Fig.~\ref{fig:StringDynamics}(a). The wave packets are initiated to move towards each other, and we probe their collision dynamics by the charge density $\langle \hat{Q} \rangle$ and electric flux $\langle \hat{E} \rangle$. In the large mass limit ($m \gtrsim \kappa$), spontaneous pair creation and annihilation are suppressed, and the dynamics of (anti)particles follow Hamiltonian \eqref{eq:eff}. Consequently, it is energetically unfavorable for the particle and the antiparticle to annihilate each other and the resulting collision is elastic. We first show this elastic collision for the non-confining case, see Fig.~\ref{fig:StringDynamics}(b)-(d). The particle and antiparticle undergo a head-on collision from $0.1$~s to $0.2$~s, and subsequently recoil in opposite directions at constant velocities. With linear fits, we find the post-collision velocities to be $v_\text{P}^\text{(f)}=-47.4(5) ~\text{sites}/\text{s}$ and $v_\text{A}^\text{(f)}=48(1)~\text{sites}/\text{s}$ for the particle and antiparticle respectively. Compared to the group velocity initiated in Fig.~\ref{fig:MovingWavePacket}(b), we find $v_\text{P}^\text{(f)} = v_\text{A}^\text{(i)}$ and $v_\text{A}^\text{(f)} = v_\text{P}^\text{(i)}$. Since the particle and the antiparticle are identical in mass, this indicates that they exchange momenta during an elastic collision, see Fig.~\ref{fig:StringDynamics}(h).

\begin{figure}[htb!]
	\centering
	\includegraphics[width=\linewidth]{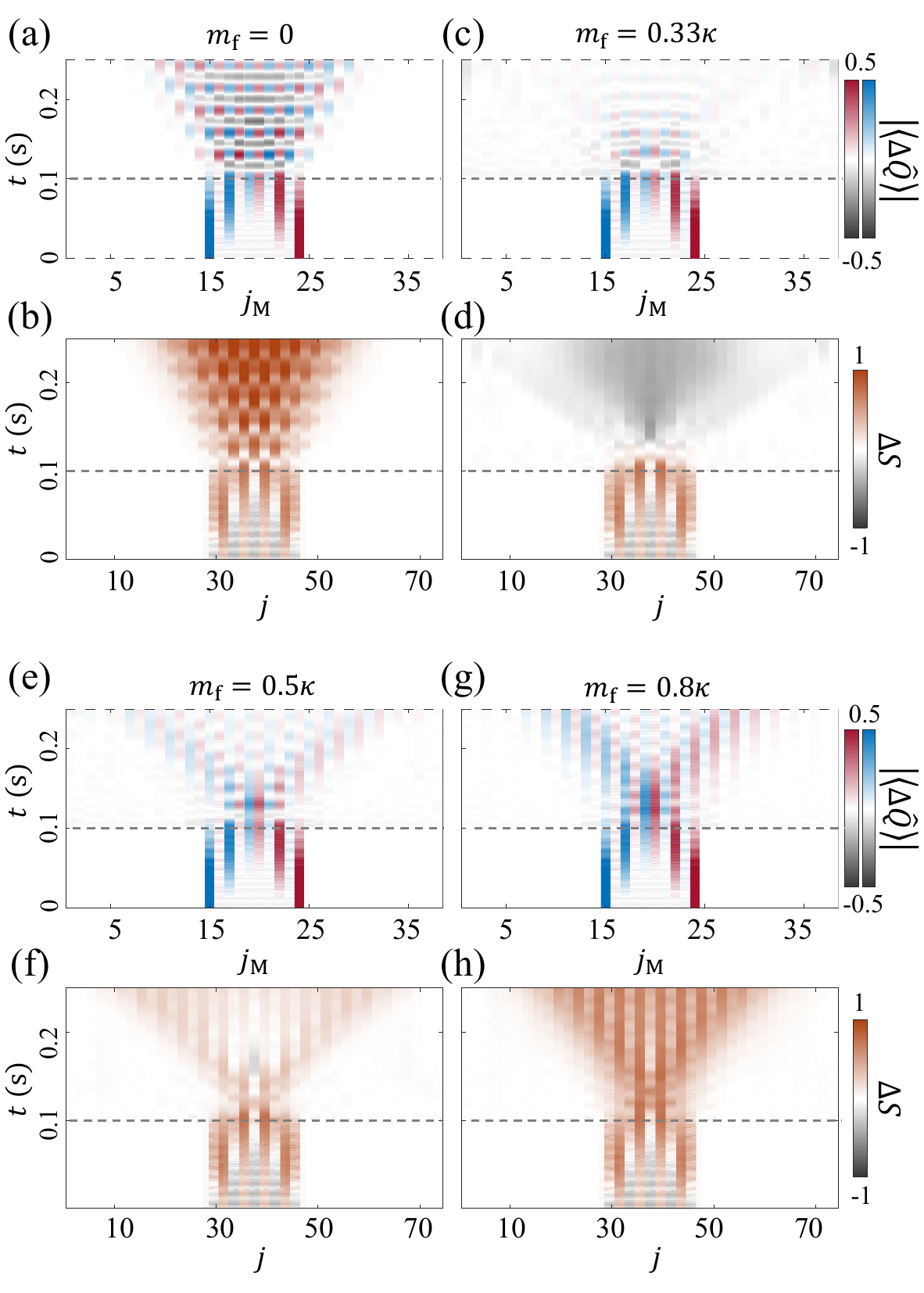}
	\caption{(Color online).
        \textbf{Quenching mass in particle-antiparticle collisions.}
        After same initialization in Fig.~\ref{fig:StringDynamics} at $m_0=1.5\kappa$, we quench to $m_\text{f}<m_0$ at $t=0.1$s. The quenches lead to spontaneous pair production in the vacuum background. For better comparison, we subtract the evolution of the pure vacuum background from the collision simulations for the same quenches and show their differences in particle density $\Delta \hat{Q}=\hat{Q}_\text{pair}-\hat{Q}_\text{vac}$ and entanglement entropy $\Delta S=S_\text{pair}-S_\text{vac}$. 
        (a) and (b): When quenching to $m_\text{f}=0$, we find that the wave packets tunnel through each other periodically, as a result of string inversions. The vacuum background undergoes scarred dynamics that deters the growth of entanglement entropy, and the colliding wave packets produce higher entropy than the scarred vacuum background. 
        (c) and (d): Around the critical point $m_\text{f}=m_\text{c}$, the vacuum background thermalizes, while the colliding wave packets oscillate at the collision point and exhibit slowed growth of entropy that leads to the negativity in panel (d).
        (e)-(h): Above the critical point, we see a suppression of the tunneling of the particle and antiparticle through each other with the suppression of pair production. Subsequently, we recover the low-energy elastic collision dynamics in Fig.~\ref{fig:StringDynamics}.
        }
	\label{fig:EPquenchmass}
\end{figure}

In the confined case with positive $\chi$, the electric flux $\ket{\triangleleft}$ has higher energy than $\ket{\triangleright}$, creating a confining force that accelerates the particle and the antiparticle towards each other. We keep the value of $\chi$ small to minimize the lattice effect (i.e. the Bloch oscillations) and therefore work within the regime of the positive effective mass ($2aq \in [-\pi/2, \pi/2]$). After the collision, the particle and antiparticle exchange momentum and recoil away from each other. The string energy increases with the inter-particle distance $D$ as they move apart, which causes deceleration of the particle and antiparticle. After reaching zero velocity, they start accelerating toward each other again, leading to the next collision. The string dynamics form a particle-antiparticle bound state oscillating dynamically in the vacuum background, which can be considered a meson, see Fig.~\ref{fig:StringDynamics}(e)-(g).

Moving on from the previous low-energy particle collisions, we bring the system out of equilibrium by an abrupt global quench of the rest mass from $m_0=1.5\kappa$ to $m_\text{f}$ at $t=0.1$~s, and thus access collision dynamics on a higher energy scale \cite{Zhou2022,Yao2022,Wang2022}. 

The vacuum background itself is unstable under the violent quenches of $m$ \cite{Zhou2022, Surace2020}. Around $m_\text{f}=0$, the vacuum background undergoes persistent oscillation between the two degenerate vacua, being an instance of quantum many-body scarring dynamics which deters the growth of entanglement entropy \cite{Turner2018, Turner2018b, Su2022, Surace2020}. Approaching the critical mass $m_\text{f} \sim m_\text{c}$, the scarring dynamics goes away and the vacuum background thermalizes with entanglement entropy maximized. When the mass increases further to $m_\text{f} \gg m_\text{c}$, the pure vacuum background is close to the ground state of the quantum link model \eqref{eq:qlm}, and entropy growth is therefore suppressed again \cite{Yao2022}. 

For $m_\text{f} \lesssim m_\text{c}$, the particle production in the vacuum background makes it difficult to distinguish the initial colliding particle-antiparticle pair, see Fig.~\ref{fig:EPquenchOriginal}. Therefore in Fig.~\ref{fig:EPquenchmass}, to better demonstrate the dynamics of collisions, we subtract the evolution of collisions with the evolution of the vacuum background (Fig.~\ref{fig:EPquenchBKG}) for the same quench parameter, and show the difference in charge density $\Delta \hat{Q}=\hat{Q}_\text{pair}-\hat{Q}_\text{vac}$ and the bipartite von Neumann entropy $\Delta S=S_j^\text{pair}-S_j^\text{vac}$, where $S_j=-\text{Tr}[\hat{\rho}_j(t) \text{ln}\hat{\rho}_j(t)]$ with the reduced density matrix $\hat{\rho}_j(t)=\text{Tr}_{k>j}\ket{\Psi(t)}\bra{\Psi(t)}$.

\begin{figure}
	\centering
	\includegraphics[width=\linewidth]{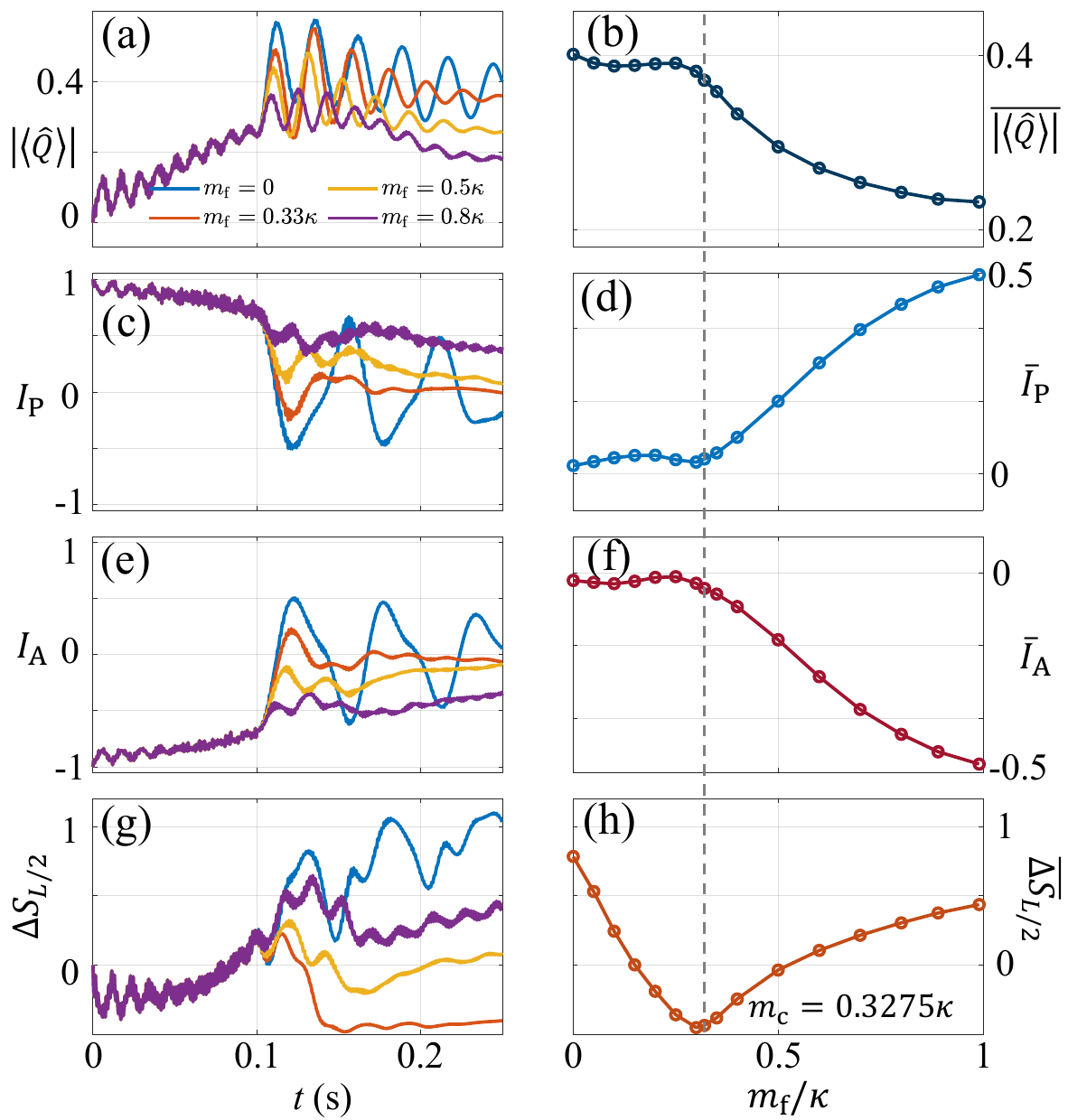}
	\caption{(Color online).
        \textbf{Quantum criticality in the particle-antiparticle collision.}
        (a) Real-time dynamics of the average charge density around the center ($\ell \in [16,23]$, see Fig.~\ref{fig:EPquenchOriginal}(c)). After the quench at $t=0.1$~s, we observe oscillations due to pair production and annihilation. 
        (b) Time-averaged charge density for different $m_\text{f}$ ($t>0.2$~s). Above $m_\text{c}$, pair production is exponentially suppressed.
        (c) and (e): Density imbalance between the left and right parts of the system for particle and antiparticle respectively. For $m_\text{f}<m_\text{c}$ the imbalance oscillates across the zero-point, indicating the particle and antiparticle tunnel through each other by string inversion.
        (d) and (f): Time average of (c) and (e). Above $m_\text{c}$, the (anti)particle is restricted to one side of the system, resulting in a non-zero imbalance.
        (g) Real-time dynamics of the difference in half-chain entropy $\Delta S_{j/2}$.
        (h): Time average of (g). Around the critical point, the colliding wave packets produce less entropy than the vacuum background.
        }
	\label{fig:QuantumCriticality}
\end{figure}

In Fig.~\ref{fig:EPquenchmass}(a), we show the quench to $m_\text{f}=0$. The colliding particle and antiparticle first annihilate each other, therefore the difference in charge density turns negative, $\langle \Delta \hat{Q} \rangle<0$. Afterward, the particle and antiparticle re-emerge due to pair production, but instead of re-emerging in their original position, the particle shows up on the right and the antiparticle shows up on the left, and in the next period they reverse in relative position again. We attribute this phenomenon to string inversion dynamics \cite{Surace2020}, where the particle and antiparticle go through each other repeatedly in the small mass limit. In Fig.~\ref{fig:QuantumCriticality}(c) and (e), we characterize the inversion by charge density imbalance between the left and right parts of the system, i.e., $I_\text{P,A}=|\langle \hat{Q} \rangle|_\text{P,A}^\text{left}-|\langle \hat{Q} \rangle|_\text{P,A}^\text{right}$. The region taken into account for the imbalance of particle $I_\text{P}$ is illustrated by the dashed blue boxes in Fig.~\ref{fig:EPquenchOriginal}(a) of Appendix~\ref{ap:detailEP}. The particle is prepared on the left with $I_\text{P}=1$ at $t=0$, after the quench $I_\text{P}$ turns negative, indicating the particle wave packet ``tunneled through'' the antiparticle to the right. The imbalance of the antiparticle $I_\text{A}$ mirrors that of the particle, as they reverse in position at the same frequency.

Around the critical point, the persistent string inversion is reduced and the vacuum background quickly thermalizes, see Fig.~\ref{fig:EPquenchBKG} in Appendix~\ref{ap:detailEP}. However, with colliding wave packets, the charge density exhibits a slower decay of oscillations, see Fig.~\ref{fig:EPquenchOriginal}(c). In Fig.~\ref{fig:QuantumCriticality}(a), we take the absolute value of average charge density near the center where particles collide, as illustrated in the dashed black box in Fig.~\ref{fig:EPquenchOriginal}(c). We fit the oscillations of the case at the critical point $m_\text{f} \approx m_\text{c}$ (orange) to a damped sine function and found the decay time to be $\tau \approx 0.065~\text{s}$, which is $2.5$ times longer than the decay time of the vacuum background ($0.026~\text{s}$ in Fig.~\ref{fig:BackGroundDynamics}(a)). 
These oscillations lead to slower growth of entanglement entropy, as illustrated by the difference $\Delta S$ (Fig.~\ref{fig:EPquenchmass}(d)). The colliding particle-antiparticle pair has a lower entanglement entropy than the vacuum background, indicating that they deter the onset of thermalization. 
Indeed, we plot the difference of half-chain entropy $\Delta S_{L/2}$ in Fig.~\ref{fig:QuantumCriticality}(g) and find distinct dynamics for different $m_\text{f}$. When we take the late-time average of them and plot with respect to the final mass, we find a dip at the critical point $m_\text{c}$. 
Microscopically, by looking at the particle density difference in Fig.~\ref{fig:EPquenchmass}(c), we see the colliding wave packets oscillate at the collision point like a metastable state, compared to the fast decaying dynamics of the vacuum background, see also Fig.~\ref{fig:EPquenchBKG}(c).

\begin{figure*}
	\centering
	\includegraphics[width=0.9\linewidth]{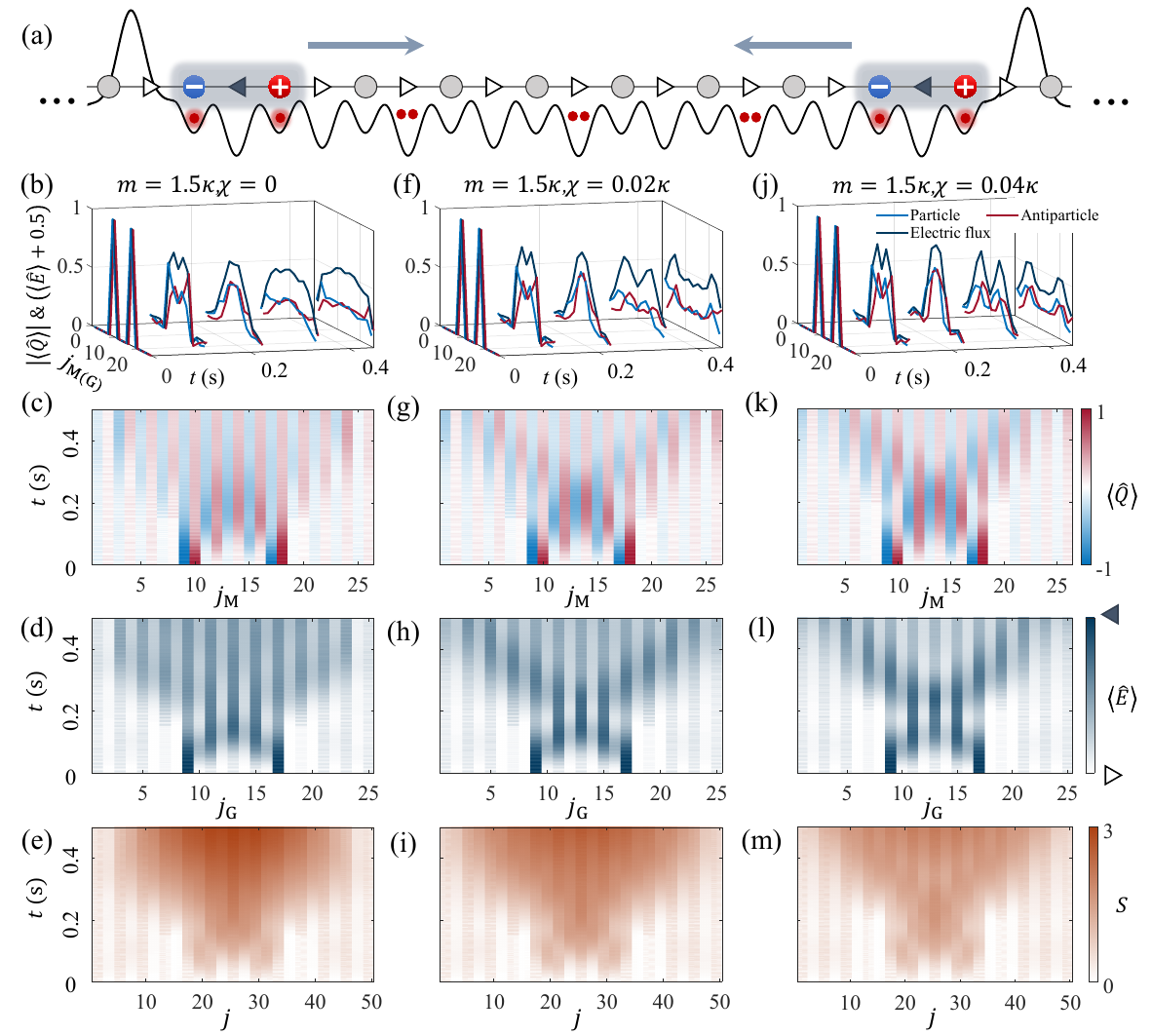}
	\caption{(Color online).
        \textbf{Meson collisions.}
        (a) A schematic illustration of meson collisions in the quantum simulator. 
        (b)-(e): Meson scattering in the deconfined case. The meson wave packet is unstable under the collision, we observe the post-collision delocalization of all wave packets. 
        (f)-(i): Meson scattering in the presence of a confining potential $\chi=0.02\kappa$. 
        (j)-(m): Meson scattering in the presence of a confining potential $\chi=0.04\kappa$.  
        As the confining potential increases, the meson wave packets become more stable. This is reflected in both the post-collision entanglement production which decreases with $\chi$, and the electric flux at the center $\langle \hat{E} \rangle _{L/2,L/2+1}$ which decreases towards the eigenvalue of the background vacuum $\ket{\triangleright}$, indicating the mesons wave packets remain localized after the collision.
        }
	\label{fig:MesonScattering}
\end{figure*}

\begin{figure}
	\centering
	\includegraphics[width=0.9\linewidth]{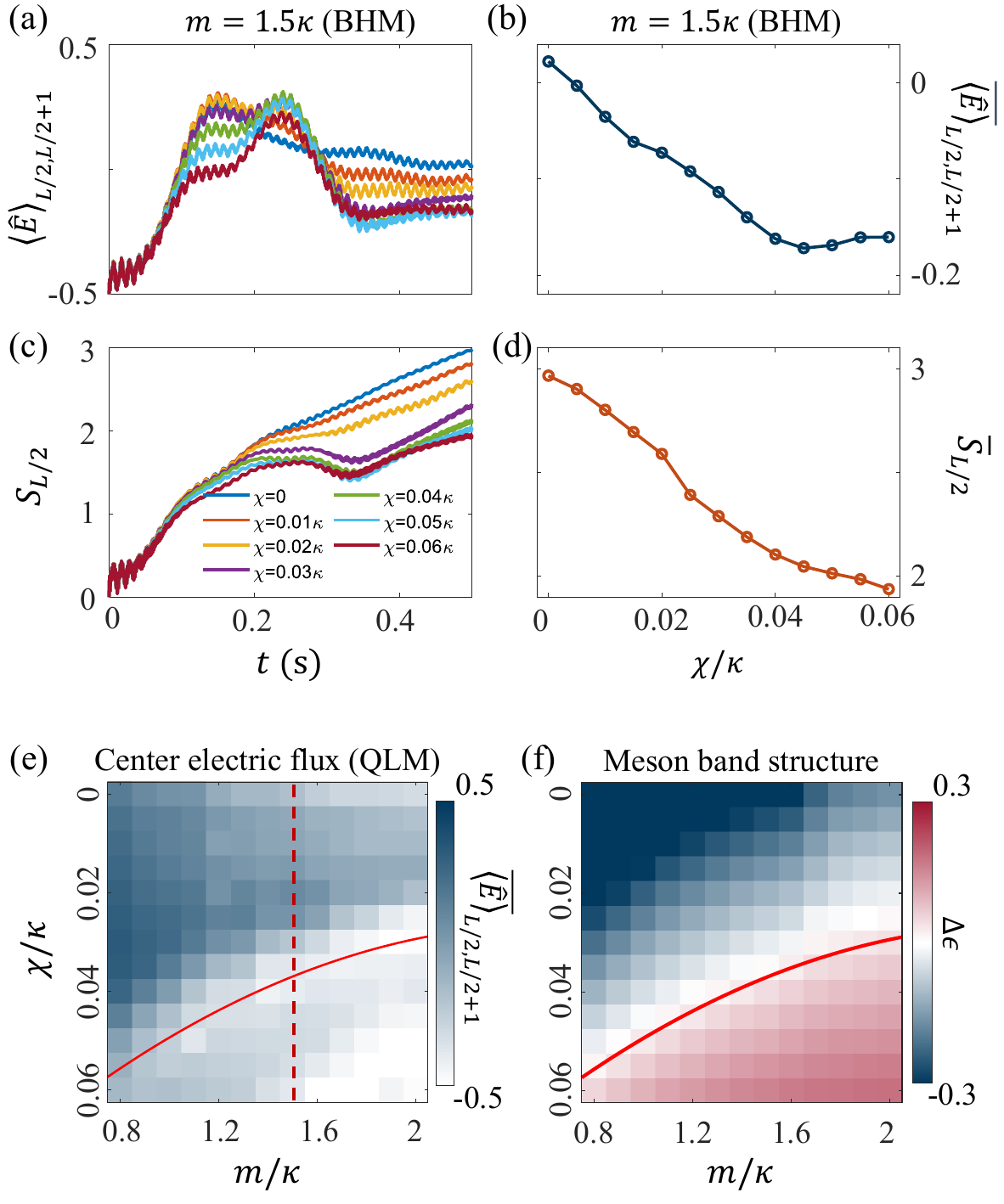}
	\caption{(Color online).
        \textbf{Probing meson band structure in the collision dynamics.}
        (a) Real-time dynamics of the center electric flux $\langle \hat{E} \rangle _{L/2,L/2+1}$, which relaxes to a stationary value after the collision.
        (b) Stationary value of (a) extracted at late times. Around $\chi=0.04\kappa$, the stationary value reaches a plateau, indicating the meson wave packets become more stable under the collision around this point.
        (c) and (d): Real-time dynamics and late-time value of the half-chain entropy $S_{L/2}$. The half-chain entropy decreases with $\chi$, as well-bounded mesons produce less entropy during the collision. 
        (e) A dynamical phase diagram showing the stationary value of center electric flux $\overline{\langle \hat{E} \rangle} _{L/2,L/2+1}$~scanning over $m$ and $\chi$, calculated in the QLM. The red dashed line points to $m=1.5\kappa$, which we calculated with the BHM in panel (b), where we find similar behavior of plateauing around $\chi=0.04\kappa$. As the mass $m$ increases, the value of $\chi$ required to reach the plateau decreases (red solid curve).
        (f) Energy difference between the first two bands of the meson and the initial energy $\Delta \epsilon = (\epsilon_1 + \epsilon_2) - (4m+2\chi) $.
        }
	\label{fig:CenterElectricField}
\end{figure}

\begin{figure}[t!]
	\centering
	\includegraphics{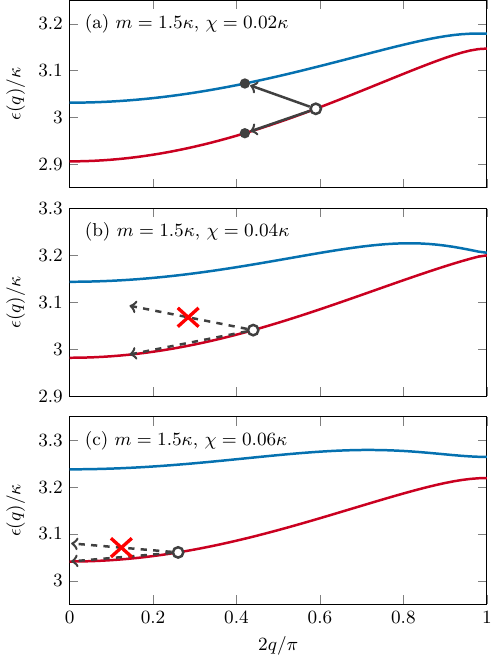}
	\caption{(Color online). \textbf{Meson band structure.}
        The excitation spectrum of the two lowest meson bands in the spin-1/2 quantum link model QLM~\eqref{eq:qlm} with lattice spacing \(a = 1\) and \(m = 1.5\kappa\), for different values of the confining potential \(\chi\), calculated numerically using the matrix product state excitation ansatz (see Appendix~\ref{sec:meson-excitations}).
        The empty circles show the initial wave-packet energies (\(2m+\chi\)), created as described in Sec.~\ref{sec:ParticleAccelerator}.
        For \(\chi = 0.02\kappa\) (a), two colliding meson wave packets can exchange relative momentum, scattering into the two states shown in the filled circles.
        For \(\chi = 0.04\kappa\) (b) and \(0.06\kappa\) (c), there is no way to scatter into the second band while conserving total energy.
        }
	\label{fig:meson-spectra}
\end{figure}

When the final mass is increased above the critical mass, particle production is exponentially suppressed, see Fig.~\ref{fig:QuantumCriticality}(b). The string inversion is thus suppressed, and the colliding wave packets can no longer tunnel through each other, see Fig.~\ref{fig:QuantumCriticality}(c) and (e). Towards $m_\text{f} \gg m_\text{c}$, we recover the low-energy elastic collision demonstrated in Fig.~\ref{fig:StringDynamics}, see also Fig.~\ref{fig:EPquenchmass}(e) and (g). As a result, the particle and antiparticle are restricted to their initial side, and the late-time density imbalance $\bar{I}_\text{P}$ and $\bar{I}_\text{A}$ become non-zero, while their absolute value increases with $m$, see Fig.~\ref{fig:QuantumCriticality}(d) and (f).

\subsection{Meson--meson collision}

We now turn to the collision of composite particles (mesons) and demonstrate how collision dynamics reveals their band structure. We focus on the large mass case with $m=1.5\kappa$ where spontaneous pair creation in the background is negligible. Following the protocol described in Sec.~\ref{sec:ParticleAccelerator}, we initiate two meson wave packets moving towards each other, see Fig.~\ref{fig:MesonScattering}(a). The barriers used to prepare the moving wave packets are removed after up to $0.2$~s. Because the mesons move faster for $\chi\leq0.02\kappa$, we remove the barriers earlier (at $0.15$~s) to avoid multiple reflections on the barrier. 

In the deconfined case ($\chi=0$), shown in Fig.~\ref{fig:MesonScattering}(b)-(e), the elementary particles and antiparticles that make up the mesons scatter elastically with no string tension between one another. We find the delocalization of all wave packets and strong entropy production after the collision. The initially localized electric flux $\ket{\triangleleft}$ spreads out throughout the whole system, indicating the breaking up of the particle-antiparticle pairs. We notice there is a refocus of wave packets at late times near the boundary, which is caused by reflections on the boundary. 

As the confining potential is increased to $\chi=0.02\kappa$, the mesons become more stable under the collision, see Fig.~\ref{fig:MesonScattering}(f)-(i). In this case, the particle and antiparticle wave packets remain localized after the collision, and their relative position remains unchanged since the particle and antiparticle can not tunnel through each other in the large mass limit. The electric fluxes $\ket{\triangleleft}$ move together with the colliding particle-antiparticle pairs, and they remain largely localized after the collision. We notice that the electric fluxes do leave a residue of $\ket{\triangleleft}$ around the center where the mesons collide, which can be observed by the electric flux at the center $\langle \hat{E} \rangle _{L/2,L/2+1}$. This residue is reduced as we increase $\chi$ to $0.04\kappa$ in Fig.~\ref{fig:MesonScattering}(l). In Fig.~\ref{fig:MesonScattering}(e), (i), and (m), we find the entropy production decreases continuously with stronger confining potential, see also Fig.~\ref{fig:CenterElectricField}(c,d). Both the entropy production and the electric fluxes indicate higher meson stability with increasing confining potential $\chi$.

To quantify this meson stability, in Fig.~\ref{fig:CenterElectricField}(a), we use $\langle \hat{E} \rangle _{L/2,L/2+1}$ as an observable and show its dynamics for different $\chi$. Its real-time dynamics first grows in time and peaks around $0.2$s as the mesons collide, while after the collision it reaches different stationary values depending on the confining potential. We plot the stationary values in Fig.~\ref{fig:CenterElectricField}(b) and find it plateaus around $\chi=0.04\kappa$. 

To understand this behavior, we simulate the meson collisions in the QLM and scan over $m$ and $\chi$. We plot the stationary values of the center electric field in a 2D dynamical phase diagram Fig.~\ref{fig:CenterElectricField}(e), where we see this plateauing behavior for different mass $m$. The value of $\chi$ required to reach the plateau decreases with increasing $m$ (solid red curve). 

This behavior is influenced by the band structure of the mesons.
In the presence of a confining potential \(\chi\), the continuum of particle-antiparticle states separates into discrete bands of bound mesons, where the higher energy bands are characterized by a greater separation between the particle and antiparticle.
As \(\chi\) is increased, the energy spacing between these bands increases, and the higher bands become more flat due to lattice effects (see Fig.~\ref{fig:meson-spectra} and Appendix~\ref{sec:meson-excitations}).
Initially, when \(\chi\) is slightly detuned from zero, the meson wave packets occupy only the lowest band.
However, during the collision, it is sometimes possible for the two mesons to scatter into a state with a lower relative momentum, where one meson is in the first band, and the other is in the second (Fig.~\ref{fig:meson-spectra}(a)).
But for larger \(\chi\), it is not possible to scatter into the second band while conserving total energy and momentum (Fig.~\ref{fig:meson-spectra}(b) and (c)).
Hence, scattering into the second band should only be allowed when the sum of energies of the first two bands at the minimum (\(q = 0\)) is less than the total initial energy of the two mesons (\(4m + 2\chi\)).
We plot the difference of these energies $\Delta \epsilon = (\epsilon_1 + \epsilon_2) - (4m+2\chi)$ in Fig.~\ref{fig:CenterElectricField}(f), which shows excellent agreement with the crossover in behavior of the stationary value of the electric flux after collision shown in Fig.~\ref{fig:CenterElectricField}(e) (red solid curves).

\section{Discussion and outlook}
We presented an experimental proposal to probe particle collisions in a $1+1$D QED theory with a state-of-the-art cold-atom quantum simulator. Using MPS numerical calculations, we demonstrated that moving wave packets of both elementary particles and composite particles can be created with potential barriers on our quantum simulator. We studied collision dynamics both near and far from equilibrium, showing that the tunability of the quantum simulator can be used to access a wide range of energy scales. By quenching mass close to Coleman's phase transition, we observed dynamics such as string inversion and entropy production in the particle-antiparticle collisions. Meanwhile, in low-energy elastic collisions, we tuned the topological $\theta$-angle to access both confined and deconfined phases, and observed string dynamics that lead to the dynamical formation of a meson state. We further demonstrated that the meson band structure could be probed with meson-meson collisions, opening the door to understanding the structure of composite particles with quantum simulation. 

Our study makes an important step towards the quantum simulation of particle collisions, which is a major objective of current working groups in the field \cite{Bauer_review,dimeglio_review}. As the underlying far-from-equilibrium dynamics of such processes can be highly nonperturbative, this in turn presents current quantum simulators with a true test of quantum advantage, which is a main driver of the field of quantum simulation in general.

An important next step is to explore the quantum simulation of particle collisions in higher-dimensional gauge theories \cite{osborne2022largescale} as well as higher-spin QLMs \cite{Osborne2023, Pichler2016}, where confinement due to the gauge coupling term becomes important. Our methods also enable the exploration of dynamical string breaking with propagating charges \cite{Hebenstreit2013, Hebenstreit2013simulating, Pichler2016}.

With our studies of the band structure of both the elementary particles and composite particles, our highly controllable and versatile gauge-theory quantum simulator also presents opportunities to explore Floquet engineering methods to study particle accelerations \cite{Hartmann2004}, Raman-assisted tunneling \cite{Eckardt2017, Aidelsburger2013, Leonard2023}, topological pumping \cite{Lohse2016, Minguzzi2022, Walter2023}, and to access physics of mesons in higher lattice bands.

Although our investigation here focuses on a one-dimensional gauge theory, the protocol of creating the moving wave packets and studying collision dynamics can be generalized to various platforms and used to study particle collisions in the Bose(Fermi)--Hubbard model, or domain wall collisions in various quantum spin models \cite{Bernien2017, Wei2022, Tan2021}, and even more exotic forms of matter such as anyons \cite{Kwan2023}.

\begin{acknowledgments}
The authors acknowledge stimulating discussions with Monika Aidelsburger, Debasish Banerjee, Yahui Chai, Philipp Hauke, Karl Jansen, Wyatt Kirkby, Ian P. McCulloch, Duncan O'Dell, Bing Yang, Zhen-Sheng Yuan, and Wei-Yong Zhang. 
This work is supported
by the Emmy Noether Programme of the German Research Foundation (DFG) under grant no.~HA 8206/1-1. This work is part of the activities of the Quantum Computing for High-Energy Physics (QC4HEP) working group.
A part of the numerical time-evolution simulations were performed on The University of Queensland's School of Mathematics and Physics Core Computing Facility \texttt{getafix}.
\end{acknowledgments}

\begin{figure}[t!]
	\centering
	\includegraphics{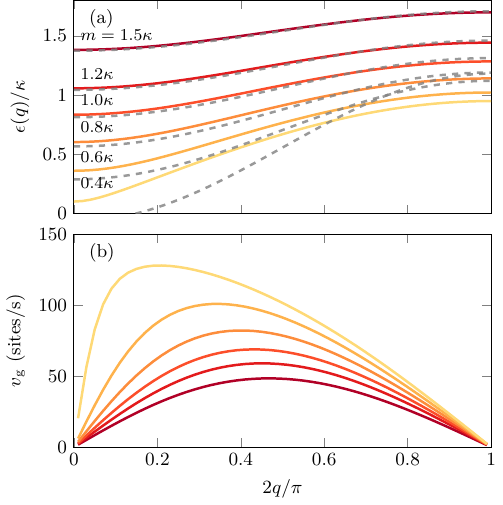}
	\caption{(Color online). \textbf{Single-particle spectra.}
        (a) The single-particle excitation spectrum of the spin-1/2 quantum link model QLM~\eqref{eq:qlm} with lattice spacing \(a = 1\) and confining potential \(\chi = 0\), scanning across various masses, approaching the critical point \(m_\text{c} = 0.3275\kappa\) from above, calculated numerically using the matrix product state excitation ansatz.
        For large \(m\), this dispersion relation approximately matches the sinusoidal dispersion relation~\eqref{eq:eff-dispersion} of the effective model~\eqref{eq:eff} (shown as the dashed gray curves on top of the data.
        (b) The corresponding group velocities.
        }
	\label{fig:single-particle-spectra}
\end{figure}

\begin{figure}[hbt!]
	\centering
	\includegraphics[width=\linewidth]{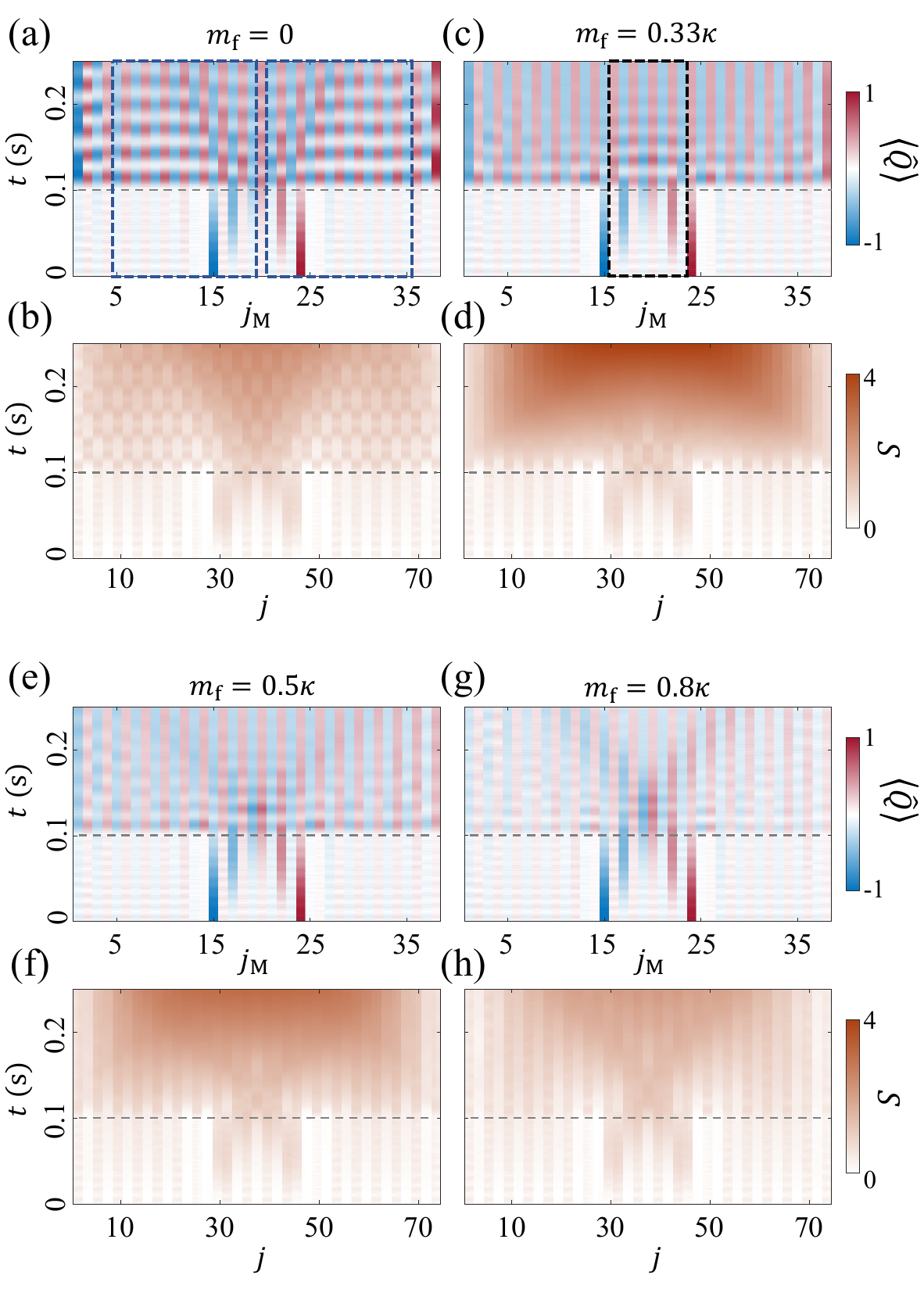}
	\caption{(Color online).
        \textbf{Particle-antiparticle collision in the wake of a mass quench.}
        (a) and (b): Quenching mass to $m_\text{f}=0$ at $0.1$~s and the scarred dynamics. The dashed blue boxes in (a) show the left and right regions used to calculate the particle imbalances $I_\text{P}$ in Fig.~\ref{fig:QuantumCriticality}(c).
        (c) and (d): A quench to the critical point $m_\text{f} \approx m_\text{c}$, where we see long-lasting oscillations in charge density (c) and a slower rise of entropy (d). The dashed black box in (c) shows the region used to calculate average charge density in Fig.~\ref{fig:QuantumCriticality}(a).
        (e)-(h): For $m_\text{f}$ above the critical point, the low-energy elastic collision dynamics in Fig.~\ref{fig:StringDynamics} is gradually recovered.
        }
	\label{fig:EPquenchOriginal}
\end{figure}

\begin{figure}[hbt!]
	\centering
	\includegraphics[width=\linewidth]{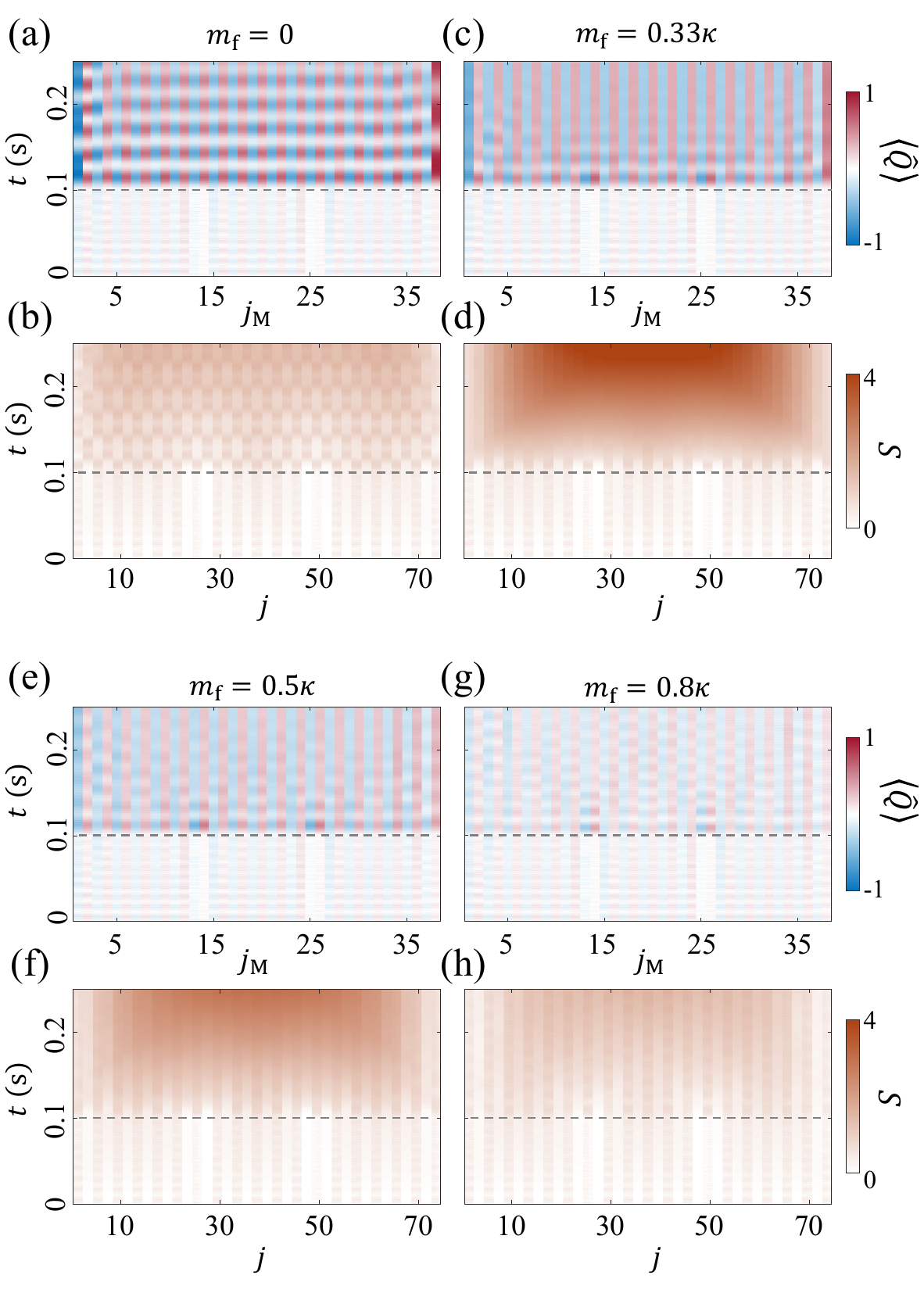}
	\caption{(Color online).
        \textbf{Evolution of the vacuum background in the wake of a mass quench.}
        (a) and (b): The background exhibits quantum many-body scarring at $m_\text{f}=0$, with slowed growth of entanglement entropy. 
        (c) and (d): Around the critical point $m_\text{c}$, the vacuum background quickly thermalizes.
        (e)-(h): Above the critical point, spontaneous pair production in the vacuum background are suppressed with the increasing mass.
        }
	\label{fig:EPquenchBKG}
\end{figure}

\begin{figure}[hbt!]
	\centering
	\includegraphics[width=\linewidth]{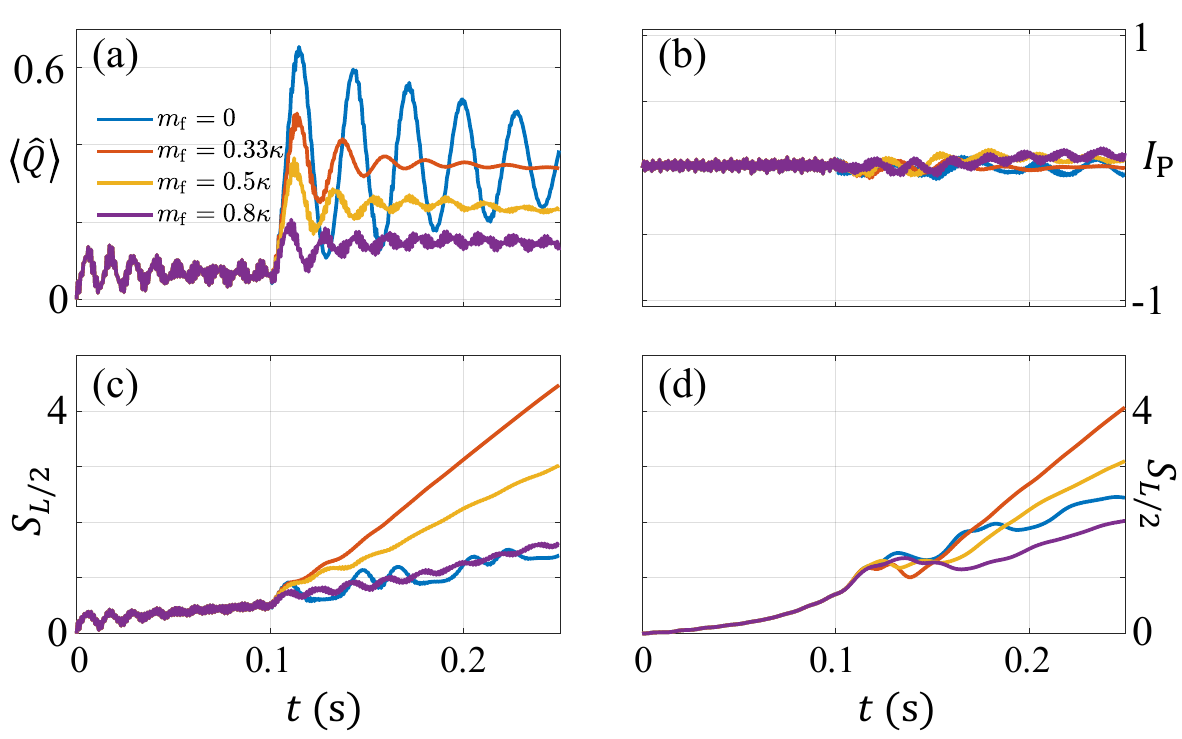}
	\caption{(Color online).
        (a) Dynamics of the average charge density of the vacuum background when quenched to $m_\text{f}$ at $0.1$~s, for the same sites as in Fig.~\ref{fig:QuantumCriticality}(a).
        (b) Left-right imbalance of the vacuum background, which shows no significant change.
        (c) and (d): Half-chain entropy of the vacuum background and the colliding wave packets for quenches to various $m_\text{f}$, respectively. Around the critical point (orange curves), the entropy of colliding wave packets exhibits an initial dip before the linear rise, which leads to lower entanglement entropy throughout the evolution times.
        }
	\label{fig:BackGroundDynamics}
\end{figure}

\appendix
\section{Perturbation theory}\label{app:PT}

The Hamiltonian for the spin-$1/2$ quantum link model (QLM) can be written in terms of the diagonal and off-diagonal terms with respect to the basis formed by the tensor product of \(\{\ket{\varnothing}, \ket{+}\}\) for even matter sites (containing antiparticles), \(\{\ket{\varnothing}, \ket{-}\}\) for odd matter sites (containing particles), and \(\{\ket{\triangleright}, \ket{\triangleleft}\}\) for gauge sites
\begin{equation}
    \hat{H}_\text{QLM} = \hat{H}_\text{off-diag} + \hat{H}_\text{diag},
\end{equation}
where
\begin{align}
    \hat{H}_\text{off-diag} &= -\frac{\kappa}{2a} \sum_\ell \left( \hat{\psi}_\ell \hat{S}^+_{\ell,\ell+1} \hat{\psi}_{\ell+1} + \text{H.c.} \right), \\
    \hat{H}_\text{diag} &= \frac{m}{2} \sum_\ell \hat{\psi}^\dagger\hat{\psi}_\ell - a\chi \sum_{\ell} (-1)^\ell \hat{S}^z_{\ell,\ell+1}.
\end{align}

If we have some initial state containing a single particle \(\ket{i} = \ket{\cdots,\triangleright,\varnothing,\triangleright,-,\triangleleft,\varnothing,\triangleleft,\varnothing,\triangleleft\cdots}\) and we wish to look at the state where the particle has jumped one particle site to the right \(\ket{j} = \ket{\cdots,\triangleright,\varnothing,\triangleright,\varnothing,\triangleright,\varnothing,\triangleright,-,\triangleleft,\cdots}\), there is no term in \(\hat{H}_\text{off-diag}\) which directly connects these two states.
However, by second-order perturbation theory, we can determine an effective Hamiltonian coupling these two states by performing a Schrieffer--Wolff transformation on the Hamiltonian
\begin{multline}
    \mel{i}{\hat{H}_\text{eff}}{j} = \delta_{ij} E_i
    + \frac{1}{2} \sum_k \left( \frac{1}{E_i-E_k} + \frac{1}{E_j-E_k} \right) \cdot \\
    \mel{i}{\hat{H}_\text{off-diag}}{k} \mel{k}{\hat{H}_\text{off-diag}}{j},
\end{multline}
where \(E_i = \mel{i}{\hat{H}_\text{diag}}{i}\).
The only state \(\ket{k}\) which will have a nonzero contribution is \(\ket{k} = \ket{\cdots,\triangleright,\varnothing,\triangleright,-,\triangleleft,+,\triangleright,-,\triangleleft,\cdots}\), and thus
\begin{align}
    \mel{i}{\hat{H}_\text{eff}}{j} &= \frac{1}{2} \left( \frac{1}{2m+a\chi} + \frac{1}{2m-a\chi} \right) \cdot \frac{\kappa}{2a} \cdot \frac{\kappa}{2a} \\
    &= \frac{m\kappa^2}{8a^2m^2-2a^4\chi^2}.
\end{align}
For \(m \gg a\chi\), we can approximate this expression by
\begin{equation}
    \mel{i}{\hat{H}_\text{eff}}{j} \approx \frac{\kappa^2}{8a^2m}.
\end{equation}
This will be the strength of the hopping term in the low-energy effective model~\eqref{eq:eff} for particles (and antiparticles). 

The approximate dispersion relation for a single particle at \(\chi = 0\) is
\begin{equation}
\label{eq:eff-dispersion}
    \epsilon(q) = m + \frac{\kappa^2}{16a^2m} \left( 1 - 4 \cos(2aq) \right),
\end{equation}
where we include an extra shift to the energy from renormalization.
The group velocity is thus
\begin{equation}
\label{eq:vg}
    v_\text{g}(q) = \frac{\kappa^2}{2am} \sin(2aq).
\end{equation}

The maximum group velocity occurs at $\sin (2aq) = 1$, which will be $v_\text{g}^\text{max} = \kappa^2/2am$.
For $aq \ll 1$, we have the linear relation
\begin{equation}
    v_\text{g}(q) \approx \frac{\kappa^2}{m} q.
\end{equation}

\section{Time evolution numerical details}

To access the dynamics of particle collisions in the Bose--Hubbard quantum simulator with minimum boundary effects, we perform numerical simulations of the BHM \eqref{eq:bhm} with a system size of $N \gtrsim 60$. We use the TEBD method implemented in the TenPy package~\cite{vidal2004, tenpy} with a time step of $1 \times 10^{-4}~\text{s} \approx 3 \times 10^{-3} J^{-1}$ and a maximum bond dimension of 3000.

For time evolution simulations of the quantum link model itself~\eqref{eq:qlm}, we use the time-dependent variational principle (TDVP) algorithm for matrix product states~\cite{Haegeman2011, Haegeman2016, mptoolkit}, using a single-site evolution scheme with adaptive bond expansion, also with a timestep of $1 \times 10^{-4}~\text{s}$.

\section{Low-lying excitation spectrum of the quantum link model}
We can calculate the low-lying excitation spectrum of the quantum link model (QLM)~\eqref{eq:qlm} using infinite matrix product state (iMPS) numerical techniques~\cite{mptoolkit}.
Specifically, we use the MPS excitation ansatz~\cite{Haegeman2012}, which is a plane-wave superposition of a local perturbation of the ground state MPS by changing a single tensor, which is then optimized with respect to energy for a specific quasi-momentum \(q\).

\subsection{Single-particle excitations}

The single-particle excitations are topologically nontrivial excitations, in that they are domain walls between the two degenerate vacua of the QLM.
(For a nonzero confining potential \(\chi\), the degeneracy between these two vacua is broken, and so this domain wall state does not have a well-defined excitation energy, so we only focus on \(\chi = 0\) here.)
In Fig.~\ref{fig:single-particle-spectra}(a) we plot the dispersion relations of the lowest-energy single particle states calculated for various \(m\).
For \(m \gtrsim \kappa\), this approximately matches the sinusoidal dispersion relation~\eqref{eq:eff-dispersion} of the effective model~\eqref{eq:eff}, but as we approach the critical point \(m_\text{c} = 0.3275\kappa\), the dispersion relation changes shape and becomes more linear around \(q = 0\).

\subsection{Bound meson excitations}
\label{sec:meson-excitations}

The meson excitations are topologically trivial excitations, that is, they are excitations on top of a single ground state.
Classically, we can picture these meson excitations as being an particle-antiparticle pair with a flux string between them corresponding to the other vacuum, which we take to be the higher energy one for \(\chi \neq 0\).
For \(\chi = 0\), the two-particle spectrum will be a continuum of scattering states, but switching to \(\chi \neq 0\) will split the low-lying spectrum into discrete oscillation modes.
As the potential energy of this flux string is linearly proportional to its length, the low-lying modes will approximately follow an Airy spectrum~\cite{Rutkevich2008}.
The higher energy modes will have a larger separation, and so the particle and antiparticle will localize apart from each other due to lattice effects, and the dispersion will be more flat~\cite{Rutkevich2008}.

In Fig.~\ref{fig:meson-spectra}, we plot the two lowest bands of the meson excitation spectrum for \(m = 1.5\kappa\) and various values of \(\chi\).
We can see that as we increase \(\chi\), the second band becomes flatter and more well-separated from the first.

\section{Details on quenches in particle-antiparticle collisions}
\label{ap:detailEP}
In Sec.~\ref{sec:epcollider} we access dynamics of various energy scales by quenching the rest mass to different $m_\text{f}$. And for a better comparison, we subtract the dynamics of the background vacuum from the colliding wave packets. Here in Fig.~\ref{fig:EPquenchOriginal}, we show the dynamics of the collision without this subtraction. With $m_\text{f} \gtrsim 0.8\kappa$ (Fig.~\ref{fig:EPquenchOriginal}(g)), the phenomenon is close to Fig.~\ref{fig:EPquenchmass}(g), as pair creation in the background is suppressed, see also Fig.~\ref{fig:EPquenchBKG}(g). When $m_\text{f}$ is reduced, the signal of initial wave packets is covered by the particles produced in the background, see Fig.~\ref{fig:EPquenchOriginal}(e). However, at $m_\text{f}=0$, the background undergoes scarred dynamics (Fig.~\ref{fig:EPquenchBKG}(a)), and the propagation of wave packets creates a phase shift with light-cone-shaped spread, similar to the observation in \cite{Surace2020}. Moreover, around the critical point, we observe longer-lasting oscillations of charge density at the collision point (Fig.~\ref{fig:EPquenchOriginal}(c)) compared to the background (Fig.~\ref{fig:EPquenchBKG}(c)). This comparison is even more clear when we compare the evolution of average charge density around the center (orange curves for $m_\text{f} \approx m_\text{c}$) in Fig.~\ref{fig:QuantumCriticality}(a) and Fig.~\ref{fig:BackGroundDynamics}(a). We extract the decay time of their oscillations by fitting to a damped sine function in Fig.~\ref{fig:BackGroundDynamics}(a), where the decay time of the orange curve is found to be $\tau \approx 0.026$~s, which is $2.5$ times faster compared to $\tau \approx 0.065$~s in Fig.~\ref{fig:QuantumCriticality}(a).  These oscillations lead to a slower thermalization, which is reflected in the half-chain entropy difference between the background (Fig.~\ref{fig:BackGroundDynamics}(c)) and the colliding wave packets (Fig.~\ref{fig:BackGroundDynamics}(d)), with their difference shown in Fig.~\ref{fig:QuantumCriticality}(g).

In Fig.~\ref{fig:BackGroundDynamics}(b) we observe that for the vacuum background, the imbalance of particle density between the left and right parts of the system shows no significant change over time after the quench, since no initial wave packets are present in the system.

\bibliography{biblio}
\end{document}